# Plasmonic Metastructures and Nanocomposites with a Narrow Transparency Window in a Broad Extinction Spectrum

*Hui Zhang* [1], *Hilmi Volkan Demir,* [2,3] *and Alexander O. Govorov* [1]*

[1] Department of Physics and Astronomy, Ohio University, Athens, Ohio 45701, USA

[2] Department of Electrical and Electronics Engineering, Department of Physics, UNAM - Institute of Materials Science and Nanotechnology, Bilkent University, Ankara 06800 Turkey

[3] School of Electrical and Electronic Engineering, School of Physical and Mathematical Sciences, Nanyang Technological University, Nanyang Avenue, Singapore 639798, Singapore

* Corresponding author: Govorov@ohiou.edu

## Abstract

We propose and describe plasmonic nanomaterials with unique optical properties. These nanostructured materials strongly attenuate light in a broad wavelength interval ranged from 400nm to 5μm but exhibit a narrow transparency window centered at a given wavelength. The main elements are nanorods and nano-crosses of variable sizes. The nanomaterial can be designed as a solution, nanocomposite film, metastructure or aerosol. The principle of the formation of the transparency window in the abroad extinction spectrum is based on the narrow lines of longitudinal plasmons of single nanorods and nanorod complexes. To realize the spectrum with a transmission window, we design a nanocomposite material as a mixture of nanorods of different sizes. Simultaneously, we exclude nanorods of certain length from the nanorod ensemble. The width of the plasmonic transparency window is determined by the intrinsic and radiative broadenings of the nanocrystal plasmons. We also describe the effect of narrowing of the transparency window with increasing the concentration of nanocrystals. Two well-established technologies can be used to fabricate such nano- and metamaterials, the colloidal synthesis and lithography. Nanocomposites proposed here can be used as optical materials and coatings for shielding of electromagnetic radiation in a wide spectral interval with a simultaneous possibility of communication using a narrow transparency window.

## Introduction

Metal nanocrystals and metamaterials with strong plasmonic resonances in the visible and infrared spectral intervals often exhibit very unique and unusual optical properties.[1,2] In single nanocrystals with specially-designed shapes, plasmonic resonance can efficiently be tuned with the geometry and can become very strong and narrow.[3,4,5] Especially there propoertyies are characteristic for the plasmonic nanorods that exhibit strong and narrow longitudinal resonances.[4] Whereas nanospheres, nanorods and nanocubes with small sizes can be grown with the colloidal synthesis in solution[3], larger-size nanostructures are conveniently fabricated by the lithographic methods.[6] Lithographically-made 2D and 3D metamaterials employ the electromagnetic interactions between building blocks to create interesting optical responses. One of the prominent effects coming from the interactions between single nanocrystals is the Fano effect. This effect can occur in purely plasmonic system[7,8,9,10] or in hybrid exciton-plasmon nanostructures.[11,12,13,14] The Fano effect typically comes from an interaction between a broad and narrow resonances in a system composed of two or several elements. Another phenomenon, which is relevant to the plasmonic Fano effect, is the plasmon-induced transparency in planar metamaterials composed of a few interacting nanocrystals.[15,16,17] Here we propose another approach to the effect of transparency window in nanostructured systems. Our approach is to construct a nanomaterial as a composition of single nanocrystals and nanocrystal complexes with narrow and tunable absorption lines.

Here we describe plasmonic media with very unusual optical properties. The idea is to design a medium or metamaterial that should attenuate light in a very broad spectral interval but exhibit a narrow transparency window at a given wavelength. Such optical materials can be used for the electromagnetic shielding that simultaneously permits communications via the transparency window. Looking at the available materials and media, we can see that the optical materials found around us may exhibit narrow absorption lines or absorption bands. For example, dye molecules have typically narrow absorption lines in the visible, or water molecules exhibit a set of narrow absorption lines in the infrared. Or, another material system widely available is amorphous carbon and carbon particles that strongly absorbs and scatter light in a wide spectral interval without any transparency intervals. A medium with the needed properties, i.e. with a broad extinction spectrum and a narrow transmission window, seems to be challenging to realize. Nevertheless, we identify theoretically the materials and designs that may lead to a realization of such media. Importantly, the media proposed in this paper are based on the real materials and realistic technologies. A narrow transparency window in a very broad extinction spectrum can exist in nanocomposites and metamaterials incorporating nanorods, nanoparticles and nano-crosses (Fig. 1). The components and composition of these optical media should be carefully designed. The elements of the nanocomposite can be simple, such as single nanorods and nanoparticles (Figs 1a and b), or more complex such as nano-crosses and multi-bar nano-crosses (Fig. 1c). The main building block of the proposed systems is a nanorod with strong, narrow and tunable extinction lines of longitudinal plasmons. The media with these unique optical spectra can be fabricated as solutions, films, coatings, planar metamaterials, etc. Nanocrystals in such media can be either randomly-oriented (Fig. 1b) or arranged in the parallel layers (Fig. 1a and c). Below we will describe the principles to design such materials as well as their optical properties.

The paper is organized in the following way. The discussion develops from the simple to the complex. The first sections concern composite media made from single plasmonic nanorods of small sizes, whereas the following sections describe media incorporating large nanorods, nano-crosses and metastructures.

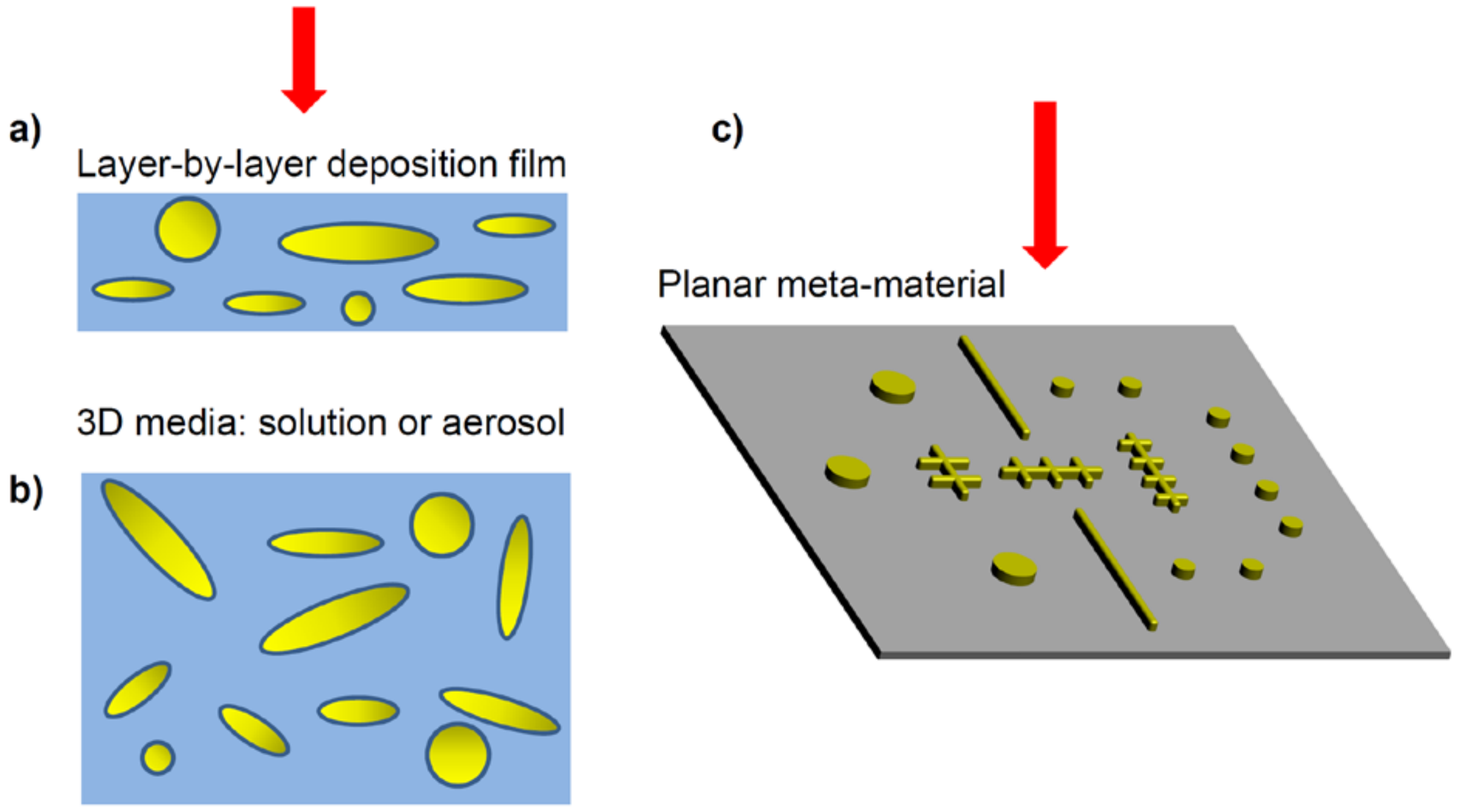

**Figure 1:** Models of the nanostructured systems discussed in the paper. **a)** and **b)**: Composite media incorporating plasmonic nanorods and nanoparticles built into a polymer film (a) and a solution/aerosol (b). **c)** Planar metamaterial composed of nanorods, nano-crosses and nano-disks. This metastructure is designed to exhibit the transparency window effect.

## 1. Small non-interacting plasmonic nanorods.

### 1.1 Optical properties of small NRs.

We start with the small plasmonic Au nanorods exhibiting narrow plasmon resonances. Optical properties of these nanocrystals are described by the analytical theory. [18,19] The extinction cross section of a small anisotropic NC is given by its absorption rate:

$$\sigma_\alpha = \frac{Q_\alpha}{I_0} = \frac{2\pi}{c_0\sqrt{\varepsilon_0}\cdot E_0^2} Q_\alpha,$$

where $Q_\alpha$ and $I_0$ are the absorption rate and the incident photon flux, respectively; $E_0$ is the amplitude of incident electromagnetic field and $\varepsilon_0$ is the dielectric constant of matrix; $\alpha$ is index describes the direction of the electric field in the incident wave and $\alpha = x, y, z$ (Figure 2). For the model of plasmonic nanorod, we choose an elongated ellipsoid with $R_x = R_y < R_z$, where $R_\alpha$ is the NR radius in the $\alpha$-direction and again $\alpha = x, y, z$. Simultaneously we adopt the notations for the length and width of NR, $a = 2R_x = 2R_y$ and $c = 2R_z$ (Fig. 2). The extinction cross section of a small ellipsoid is given by the following analytical equations:[18,19]

$$Q_{MNP,\alpha} = \mathrm{Im}(\varepsilon_{Au})\frac{\omega}{2\pi}V_{NC}\left|\frac{\varepsilon_0}{(\varepsilon_{Au}-\varepsilon_0)L_\alpha+\varepsilon_0}\right|^2,$$

$$L_x = \frac{1-e^2}{2e^3}\left(Ln\frac{1+e}{1-e}-2e\right),\quad L_y = L_z = \frac{1}{2}(1-L_x),\quad e=\sqrt{1-\left(\frac{R_z}{R_x}\right)^2}, \qquad (1)$$

where $\varepsilon_{Au}$ and $\varepsilon_0$ are the dielectric constants of nanocrystal and matrix, respectively, and $V_{NC} = (4/3)\pi R_x^2 R_z$ is the NR volume. In a 3D medium incorporating randomly-oriented NRs, the cross section should be averaged over the orientations,

$$\sigma = \frac{\sigma_x + \sigma_y + \sigma_z}{3}.$$

Sometimes it is more convenient to use, instead of the extinction cross section, a molar extinction coefficient defined as

$$\varepsilon_{[(M\cdot cm)^{-1}]} = N_A\frac{10^{-4}}{0.23}\cdot\sigma_{[cm^2]}.$$

The extinction spectrum of a single NR exhibits a strong resonance related to the longitudinal plasmon at a long wavelength and a relatively weak peak for the transverse plasmon at a shorted wavelength (Fig. 2a). To realize a material with a narrow transparency window, we will utilize the strong L-plasmon resonance. We therefore look now at the function $\sigma_z(\omega)$ and investigate the position and broadening of L-plasmon as a function of the NR aspect ratio

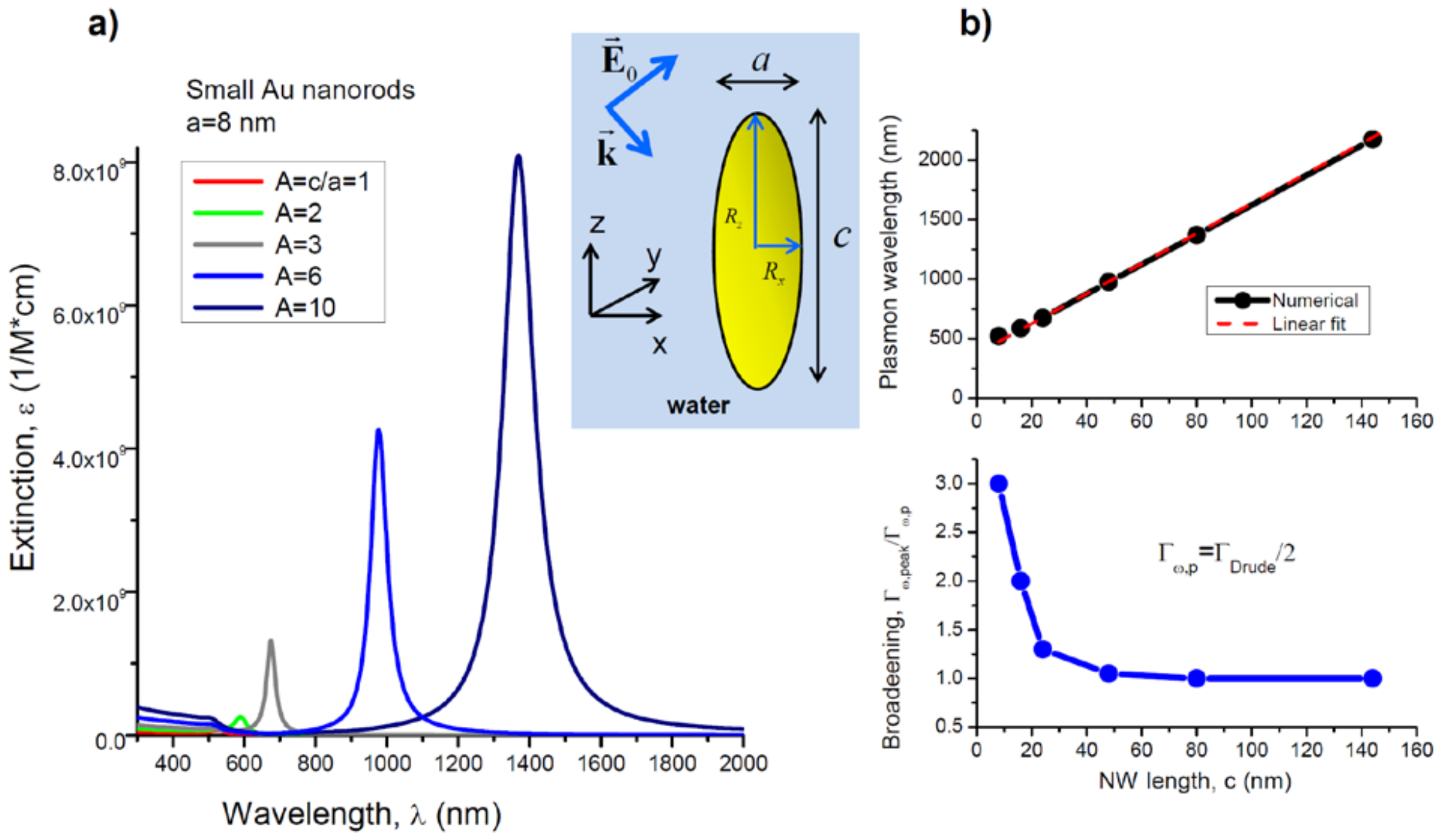

**Figure 2:** **a)** Extinction spectra of small Au nanorods for various aspect ratios. The width of NRs is kept constant, $a = 2 \cdot R_x = 8nm$. The excitations were averaged over the orientation of incident electric field. **b)** Wavelength and broadening of the L-plasmon peak as a function of the NR length.

$$A = \frac{R_z}{R_x} = \frac{c}{a}.$$

This can be done numerically using the empirical dielectric function.[20] Or we can use the Drude model,

$$\varepsilon_{Au}^{D}(\omega) = \varepsilon_{0,Au} - \frac{\omega_p^2}{\omega(\omega + i\Gamma_{Drude})},$$

that provides a good approximation for the experimental dielectric constant in the wavelength interval $\lambda > 600nm$. The parameters of the Drude model should be found from the fit to the experimental dielectric function[20] and are given by the following numbers: $\varepsilon_{0,Au} = 9.07$, $\omega_p = 8.91eV$ and $\Gamma_{Drude} = 0.076meV$. We also should mention about the inter-band absorption effects in gold in the wavelength interval $\lambda < 600nm$. These inter-band effects can be approximately described by the Lorentz-Drude model[21]

$$\varepsilon_{Au}^{LD}(\omega) = \varepsilon_{LD,Au} + \Delta\varepsilon_{L,\text{inter-band}}(\omega) - \frac{\omega_p^2}{\omega(\omega + i\Gamma_{Drude})},$$

where $\Delta\varepsilon_{L,\text{ inter-band}}(\omega)$ is the Lorentz term coming from the inter-band transitions.

The position of the L-plasmon peak in a NR is given by the pole of Eq. 1 and therefore can be found from the equation $(\varepsilon_{Au} - \varepsilon_0)L_\alpha + \varepsilon_0 = 0$. For the elongated NRs ($R_x = R_y << R_z$), we obtain the complex frequency of the L-plasmon in the form

$$\omega_{res} = \Omega_{p,L} - i\Gamma_{Drude} / 2,$$

$$\Omega_{p,L} = \frac{\omega_p}{\sqrt{\varepsilon_{Au} - \varepsilon_0 \left(1 - \frac{1}{L_z}\right)}} \approx \frac{\omega_p \sqrt{Ln(A)}}{A\sqrt{\varepsilon_0}}. \qquad (2)$$

Equation 2 was derived for the limit $A >> 1$. In particular, we see from Eq. 2 that the plasmon resonance broadening of NR stays the same in the long-wavelength spectral interval and is given by the intrinsic losses in the metal. While the plasmon-peak broadening of the cross section $\sigma_z(\omega)$ stays constant as a function of the NR length, the width of the L-plasmon resonance ( $\Gamma_{p,\lambda}$ ) of the function $\sigma(\lambda)$ grows with increasing $R_z$,

$$\Gamma_{p,\lambda} = \Gamma_{p,\omega} \frac{\lambda_{p,L}^2}{2\pi c_0} = \lambda_L \frac{\Gamma_{p,\omega}}{\Omega_{p,L}}, \qquad (3)$$

where

$$\Gamma_{p,\omega} = \frac{\Gamma_{Drude}}{2}$$

is the broadening of the L-plasmon coming from Eq. 2, $\lambda_{p,L}$ is the L-plasmon wavelength which is an increasing function of the aspect ratio. These properties of the L-plasmon peak are summarized in Fig. 2b. Numerically, the L-plasmon wavelength $\lambda_{p,L}(A)$ for the gold NRs can be approximated as a simple linear function for the aspect ratios $A > 1$:

$$\lambda_L(A) = \lambda_L^0 \left(1 + \beta \cdot A\right).$$

where $\lambda_L^0 = 374nm$ and $\beta = 0.033$. The calculated linear dependence of the L-plasmon peak is in good agreement with the experiments.[4] The magnitude of the L-plasmon peak in the absorption spectrum strongly increases with increasing NR length. By analyzing equation 1 for long ellipsoids, we can see that the peak absorption of long NR is a linear function of the NR width, $\sigma_{\text{z,peak}} \sim V_{NR} \sim R_z$. We see this behavior in Fig. 2a. For aspect ratios $A \sim 1$, the increase of $\sigma_{\text{z,peak}}$ is strongly enhanced by the narrowing effect. The narrowing effect of L-plasmon peak for the long NRs is another important feature. A general equation for the broadening of L-plasmom peak in the spectrum $\sigma(\omega)$ is given by

$$\Gamma_{peak,\omega} = \Gamma_{p,\omega} + \Delta\Gamma_{\text{inter-band}},$$

where $\Delta\Gamma_{\text{inter-band}}$ is the plasmonic broadening due to the inter-band processes in gold, which become very prominent in the interval $\lambda < 600nm$. Simultaneously the broadening $\Gamma_{p,\omega}$ comes from the intra-band scattering processes. Mathematically, the inter-band contribution $\Delta\Gamma_{\text{inter-band}}$ can be derived using the Drude-Lorentz dielectric function given above. With increasing aspect ratio, the inter-band broadening vanishes and the plasmon-resonance width becomes equal to the residual Drude broadening $\Gamma_{p,\omega} = \Gamma_{Drude}/2$ (Fig. 2b). Therefore, we observe the narrowing effect for the L-plamson resonance in long NRs. Simultaneously, we note that the L-plasmon peak width, $\Gamma_{p,\lambda}$, for the function $\sigma(\lambda)$ is an increasing function of the NR width, as we see in Fig. 2a and as it follows from Eq. 3.

**1.2 Collections of NRs.**

Our first model is a composite medium that incorporates small plasmonic NRs with random positions and orientations. We assume that the ensemble of NRs is diluted and Coulomb and electromagnetic interactions between NRs are weak and can be neglected. We now construct a medium from a mixture of NRs of various sizes according to simple rules outlined below. Designing collection of NRs, we do not include nanocrystals with a certain length and, therefore, the transmission spectrum of such composite material acquires a window centered at a certain wavelength. The best transmission windows can be made from an infinite collection of small NRs of various sizes and this collection of NRs should be represented by a continuous size-distribution function. However, practical implementations of this simple idea will require a discrete sequence of NRs and we will give below related examples.

### 1.3 Continuous collection of NRs.

We now consider a collection of NRs with a constant width $R_x = R_y$ and a variable length $R_z$. A size distribution of NRs is defined by the distribution function $\delta n_{NR}(R_z)$ that can be normalized to the total density of NRs

$$n_{tot} = \int_{R_{z,\min}}^{R_{z,\max}} dr_z \cdot \delta n_{NR}(r_z),$$

where $r_z$ is a NR z-radius that is a variable in this case. The resulting averaged cross section of a composite is given by the integral

$$\bar{\sigma} = \frac{1}{n_{tot}} \int_{R_{z,\min}}^{R_{z,\max}} dr_z \cdot \delta n_{NR}(r_z) \cdot \sigma(r_z)\,.$$

Then, the transmitted intensity and the transmission coefficient of a mixture are given by

$$I_t = I_0 e^{-\alpha},\ T = \frac{I_t}{I_0} = e^{-\alpha} = 10^{-OD}$$

where the exponents, that are the optical densities, should be calculated as

$$\alpha = L_{path} n_{tot} \bar{\sigma} \equiv L_{path} \int_{R_{z,\min[nm]}}^{R_{z,\max[nm]}} dr_z \cdot \delta n_{NR}(r_z) \cdot \sigma(r_z),$$

$$OD = \frac{\alpha}{Ln[10]},$$

where $L_{path}$ is the optical path and $\sigma(r_z)$ is the NR absorption cross section averaged over the orientations. Now we will try this approach with the following model density distribution

$$\delta n_{NR,0}(r_z) = n_0 \frac{1}{r_{z[\mathrm{nm}]}^{2.6}}\,, \qquad (4)$$

where $r_{z[\mathrm{nm}]} = c/2$ is the NR z-radius counted in nm. The other parameters are chosen as $R_{\mathrm{z,\,min[nm]}} = 8$, $R_{\mathrm{z,\,max[nm]}} = 440$ and $n_0 = 9 \cdot 10^{13} cm^{-3}$; the corresponding total density and molar concentration are calculated as $n_{tot} = 2 \cdot 10^{12} cm^{-3}$ and $c_{tot} = 3.4 \cdot 10^{-9} M$. Since we assume that the distribution of NRs is continuous and obeys exactly the above law, this is certainly a model case. At the same time, the density of the NR mixture in this case is a realistic number. Now we remove NRs with the sizes in the interval $25 < r_{z[nm]} < 35$ from this mixture and obtain the following distribution function

$$\delta n_{NR,window}(r_z) = n_0 \frac{1}{r_{z[\mathrm{nm}]}^{2.6}} \left( \theta(r_{z[\mathrm{nm}]} - 35) + \theta(25 - r_{z[\mathrm{nm}]}) \right). \qquad (5)$$

The corresponding gap in the distribution of NR length is in the interval $50nm < c < 70nm$. In Figs. 3, we illustrate the model of collection of NRs from which we remove NRs with certain sizes ($50nm < c < 70nm$). Figure 4 gives now the distribution $\delta n_{NR,window}$ and the calculated optical density for NRs with $a = 12nm$. We see an excellent transmission window in the spectrum. This sharp and well-defined transmission window appears because the individual Au NRs exhibit very narrow plasmon peaks. The transmission window takes place in the interval $780nm < \lambda < 975nm$ which corresponds to the plasmonic wavelengths of the NRs with the lengths $c = 50$ and $70nm$. This case shows the principle to create a narrow window in the transmission spectra of plasmonic composites.

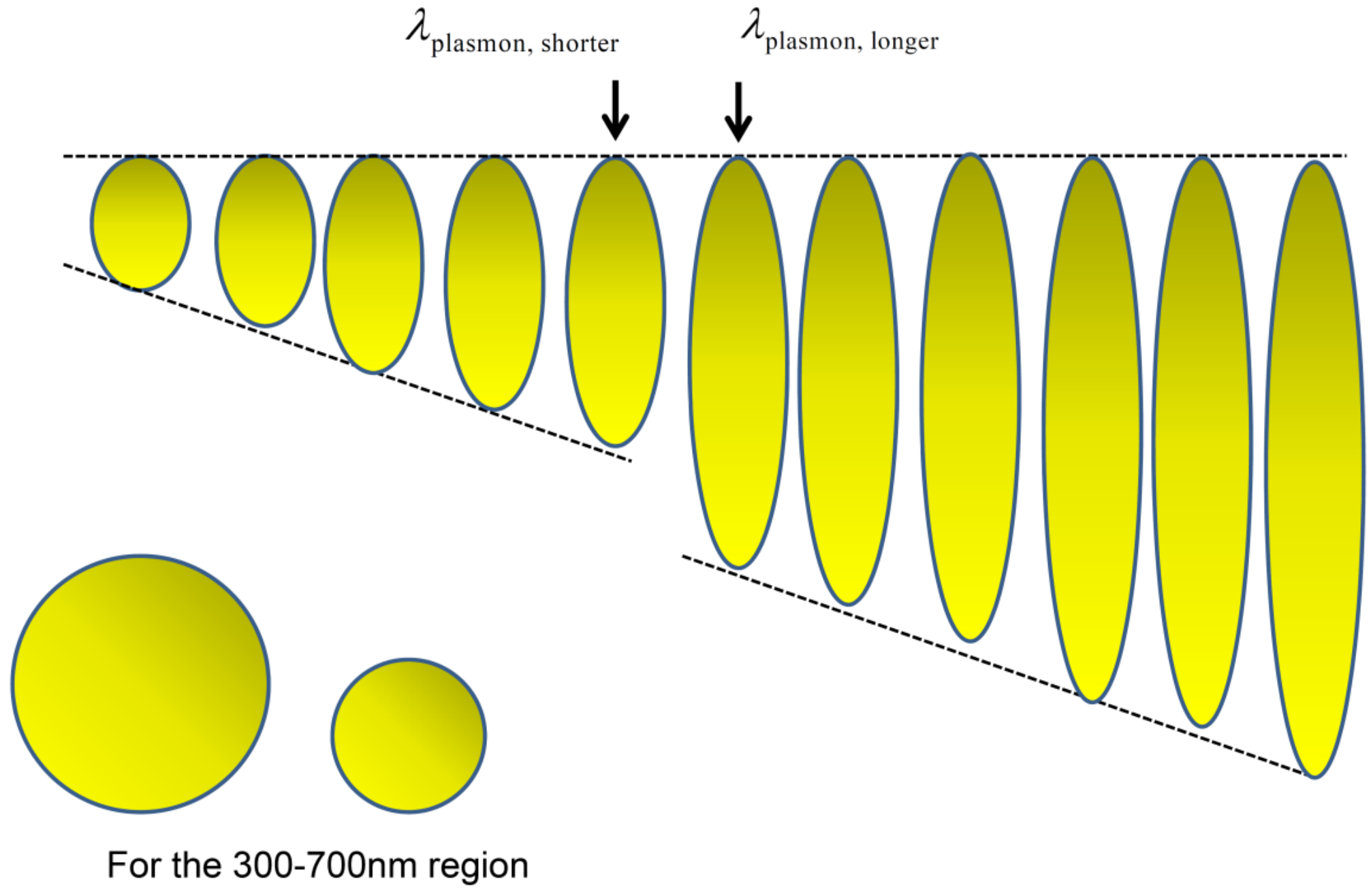

**Figure 3:** Model of collection of NRs that exhibits the transparency window effect in the transmission spectra. The NR collection does not include NRs with lengths in the interval $50nm < c < 70nm$. Correspondingly, the NRs with the gap sizes ($c = 50nm$ and $c = 70nm$) determine the positions of the edges of the transparency window $\lambda_{\text{plasmon, shorter}}$ and $\lambda_{\text{plasmon, longer}}$. The NR collection can be supplemented with spherical nanoparticles that create strong absorption in the interval 300-700nm.

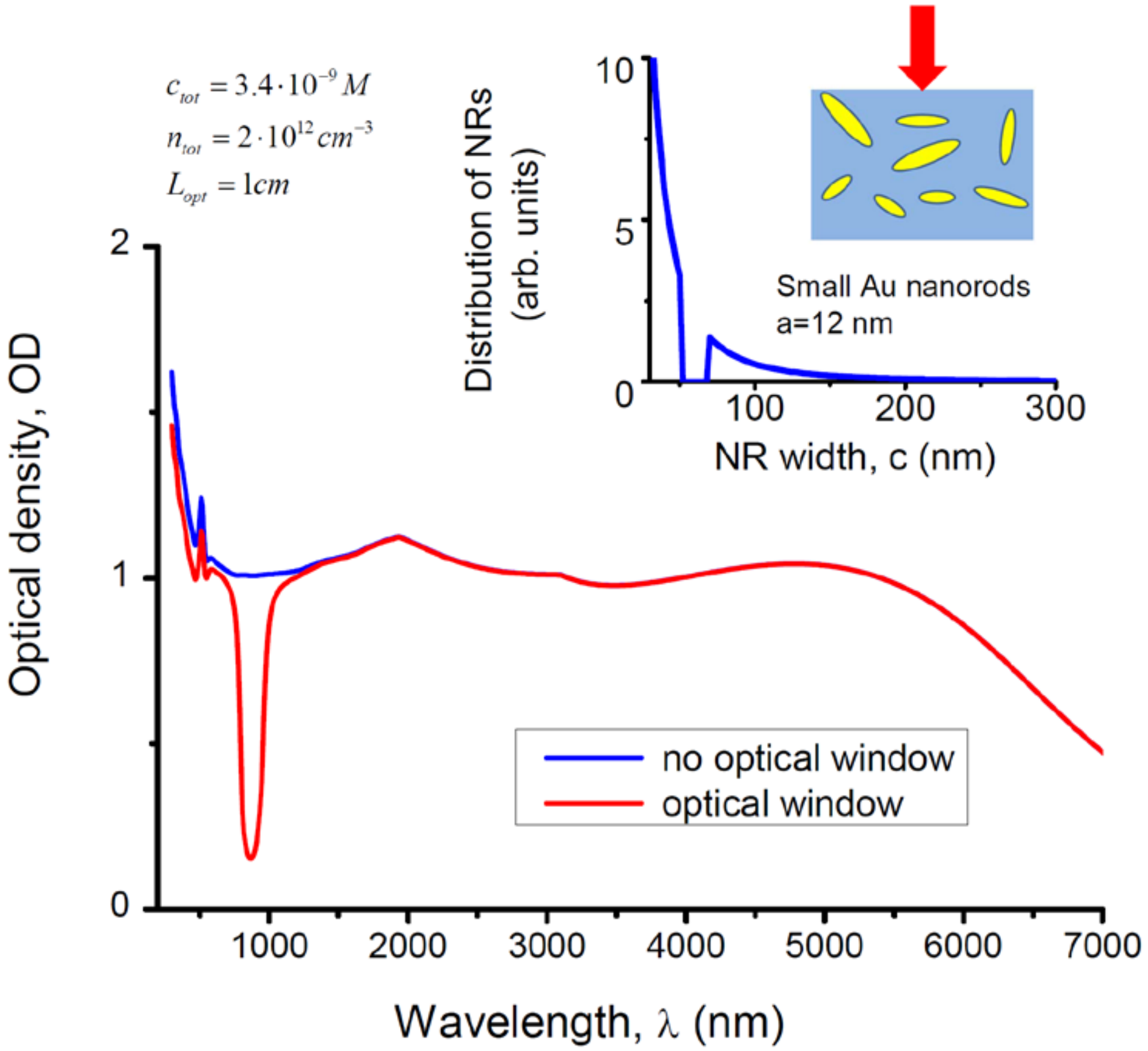

**Figure 4:** Optical density for the model collections of NRs. Blue and red curves show the absorptions for the NR collections without and with the window effect, respectively. These curves were calculated from the NR distributions given by Eqs. 4 and 5. Inset: The NR-size distribution with a gap from 50 to 70nm. The inset also illustrates a composite medium with randomly orientated NRs.

### 1.4 Discrete collection of NRs.

From the above model case, we now move to a more realistic case of mixture of NRs with a discrete distribution of sizes. Namely we take a composite of 12 kinds of NRs and 2 kinds of NPs (Figure 5). This composite gives an optical window at ~ 900nm. Spherical nanoparticles are used to "close" the absorption spectrum in the interval 400-600nm. One NP size is small, $12nm$ and another is large, $150nm$. To calculate the extinction of the large-size NPs, we used the full electromagnetic Mie theory[22] because these NPs are of relatively large sizes. The optical density of the NR-NP composite is given now by the sum:

$$OD = \frac{L_{path}}{Ln[10]} \sum_i n_i \cdot \sigma_i,$$

where the index $i$ labels the kind of nanocrystals (NR and NP) and $\sigma_i$ is the corresponding cross section. Taking a sufficiently long optical path of the NR-NP composite medium, we obtain a spectrum of transmission with a well-defined optical window at about 900nm (Figure 5). In Supporting Information we also show the contributions of individual nanocrystals to the total optical density. This particular composition was chosen to create both a wide and nearly-flat absorption spectrum in the region 400nm-1.5μm and a narrow transparency window at about 900nm. In this composition, small Au NPs can be substituted by semiconductor quantum dots (QDs) with an appropriate band gap. For example, CdTe QDs strongly absorb in the region $\lambda < 800nm$ and can be used as an element of the composite. This case is described in Supporting Information.

This theoretical demonstration of the transparency-window effect assumes a liquid solution with nanocrystals or a polymer composite film. However, the same window effect can be realized with an aerosol of nanocrystals when nanocrystals are dispersed in air. In this case, the concentrations can be reduced, but the optical path can be correspondingly increased. For example, if we take an optical path in air as *1m*, the nanocrystal densities in the table given in Fig. 5 should be divided by 100. Then, we obtain similar spectra with a transparency window.

NRs with small sizes can be fabricated with the well-established colloidal synthesis approach.[3,4] To demonstrate the transparency window effect, we used NRs with a diameter of *12nm* and lengths of *44 - 132nm*. Of course, this effect can be demonstrated with many other collections of NR sizes. The particular NR sizes used by us here are very typical for the colloidal nanocrystals. For example, Nanopartz[23] cells Au NRs with very well-defined sizes in the intervals $20nm < c < 250nm$ and $10nm < a < 50nm$. To realize the transparency window effect experimentally, one can think about a few experimental systems: (1) Colloidal NRs that can be dispersed in liquid solution; (2) NRs composition can be built into a polymer film or coating. For this, among other approaches, one can employ the layer-by-layer deposition technique; (3) Individual NRs or polymer particles containing NRs can be despised in air to form a plasmonic aerosol.

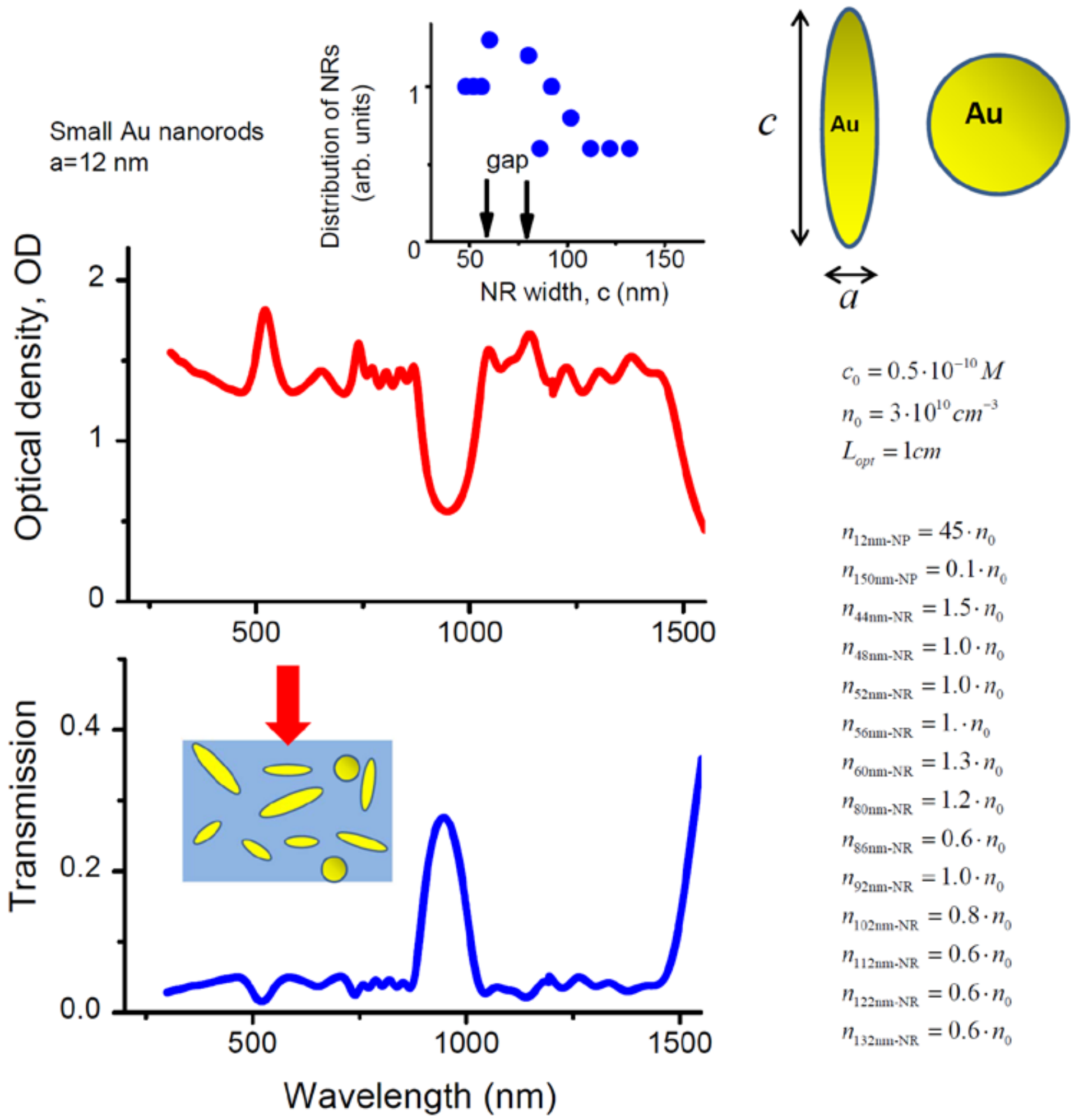

**Figure 5:** Optical density and transmission of the medium incorporating a discrete size-distribution of NRs. On the right-hand side, we list the NRs and their concentrations. This composite contains also spherical Au NPs of two sizes. Insets show the distribution of NR sizes and the model of plasmonic composite medium. A matrix for this media can be liquid or polymer.

## 2. Large non-interacting plasmonic nanorods: Electromagnetic effects.

### 2.1 Electromagnetic properties of single NRs.

Large NRs can be fabricated on a substrate using the lithography. Then, optical properties of NRs can be measured directly on a substrate, or NRs can be released to a liquid solvent[24,25,26] where the optical absorption can be recorded. As another processing step, NRs in solution can be deposited on a polymer film and then covered with another polymer layer using the standard layer-by-layer technology.

NRs with sizes comparable to the wavelength of light exhibit strong retardation effects. Such nanorods should be calculated numerically. In Figures S4 and 6 we plot the extinctions of nanorods computed numerically by the Final Elements Method (COMSOL). We choose the geometry and sizes accessible for the lithographical nano-fabrications and, therefore, the NR dimensions are taken larger than those for the colloidal NRs. Simultaneously, since we aim at the transparency window effect we choose relatively small sizes to avoid strong radiative broadening of the L-plasmon peaks. Namely, we take the NR width and height as $w = 40nm$ and $h = 50nm$. The NR length, $L$, is a variable parameter. First of all, we see that the extinction now has a strong component of scattering and this scattering contribution starts to dominate for large sizes (Fig. S4). Another important effect is the radiative broadening. With increase of NR size, the plasmon peak becomes much broader (Fig. 6). This is the effect of radiational decay of plasmon. The broadening of the plasmon peak is now composed of two terms:

$$\Gamma_{peak,\omega} = \frac{\Gamma_{Drude}}{2} + \Gamma_{radiative}.$$

The contribution $\Delta\Gamma_{\text{inter-band}}$ can be neglected since we deal with the L-plasmons with $\lambda_p > 600nm$. Another observation is that the NR geometry creates typically narrower plasmon peaks compared to the nano-disks (Fig. 6b). Therefore, to realize the optical transparency window effect, we prefer to use the NRs.

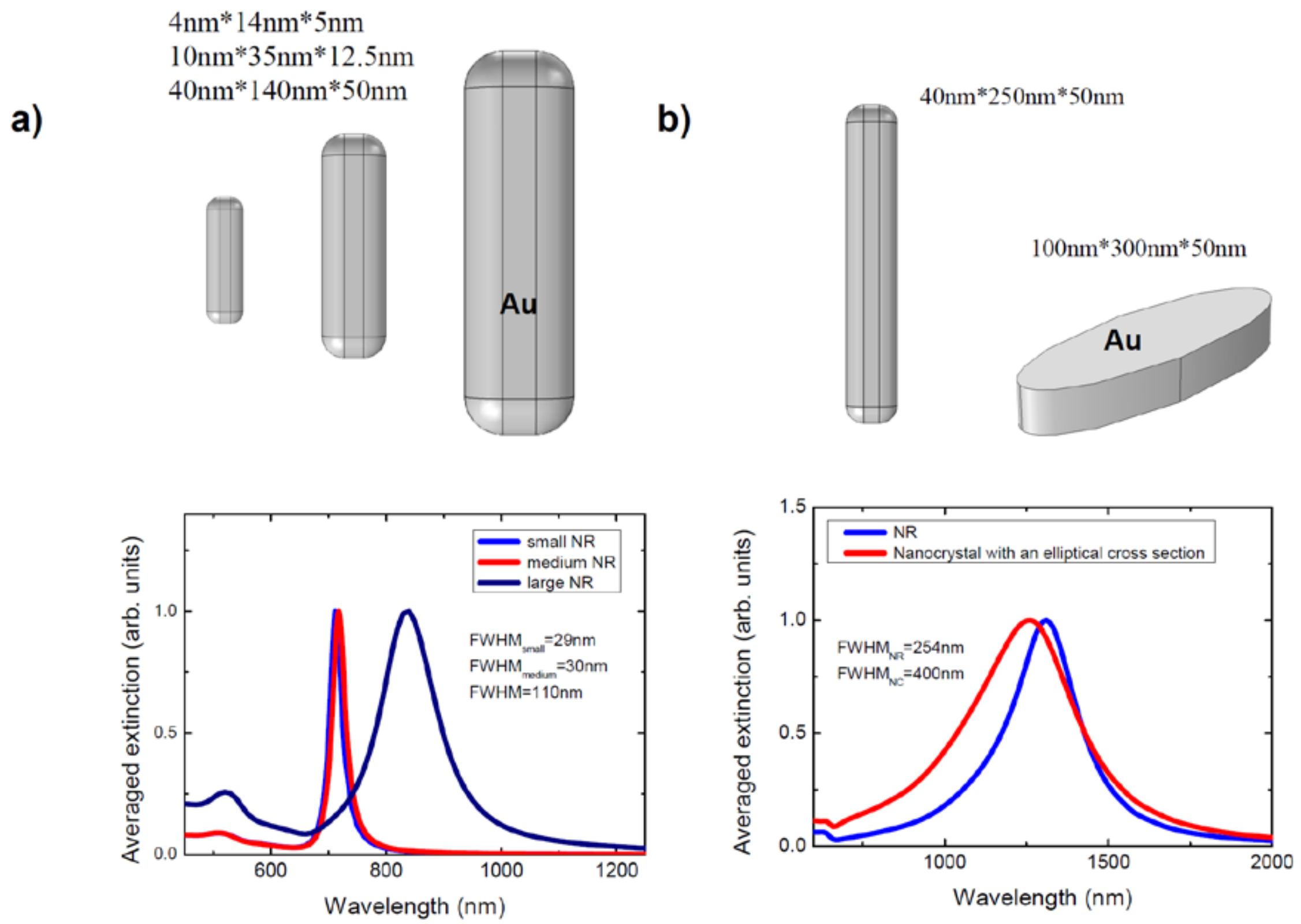

**Figure 6:** Electrodynamic properties of nanorods and nanodisks. **a)** Broadening of the L-plasmon peak strongly increases with a nanocrystal size due to the radiative decay. **b)** NRs typically exhibit more narrow L-plasmon lines compared to the case of nanodisks with a similar plasmonic wavelength.

We now write down some convenient analytical equations for large Au NRs. The plasmon wavelength and the intensity of the plasmon peak increase with increasing length of NR. For NRs with $w = 40nm$ and $h = 50nm$, the following formulas can be used as a good fit to the numerical data

$$\lambda_{p[nm]} = 1070nm + 4.2(L_{nm} - 200nm),$$
$$\sigma_{[m^2]} = 4.8 \cdot 10^{-20} L_{[nm]}^{1.45} \cdot \frac{\lambda_{p[nm]}^{2.4}}{(\lambda - \lambda_{p[nm]})^2 + 4.3 \cdot 10^{-4} \lambda_{p[nm]}^{2.4}}.$$

From this formula, we see that the broadening of the plasmon peak increases with the NR length, $\Gamma_{\lambda,nm} \approx 0.02 \cdot \lambda_{p,nm}^{1.2}$.

**2.2 Discrete collection of large NRs.**

Large plasmonic NRs, which can be fabricated lithographically, allow us to construct plasmonic media with a narrow transparency window in a much wider wavelength interval. In principle, one can construct a material with a transparency window located in the interval from *700nm* to *1mm*. We now describe one realization of the NR composition with the optical window around 1500nm. For this, we choose NRs in the interval *125-900nm* and again create a gap in the size distribution. In this case, the gap is taken between *250* and *420nm*. It means that the NRs with the L-plasmon wavelengths around *1500nm* are not present in the collection. As before, we supplement the NR collection with large spherical Au NPs that create strong absorption for $\lambda \leq 600nm$. Figures 7 and S5 show the calculated spectra.

One interesting observation in the calculated transmission spectra is that the transparency window becomes narrower with an increase of optical path because of the non-linear property of the exponential function (Figures 7 and S5). The same effect can be achieved if we increase the concentration of particles. Therefore, the window effect can be enhanced either with an increase of concentration of nanocrystals or by making an optical path longer.

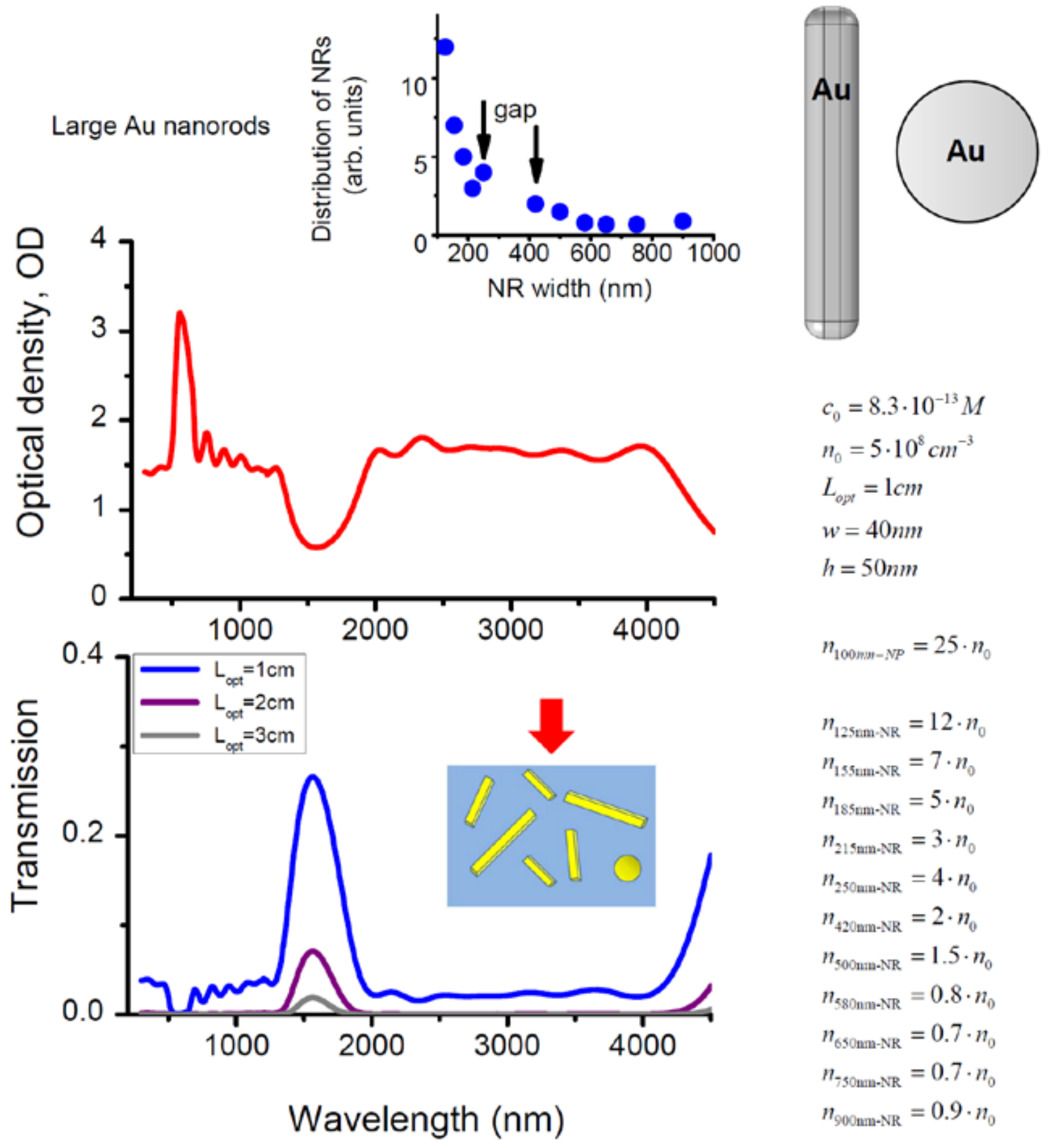

**Figure 7:** Optical density and transmission of the medium incorporating a collection of large NRs and NPs. On the right-hand side, we list the concentrations of nanocrystals. Insets: Distribution of NR sizes and the model of plasmonic composite medium incorporating both NRs and NPs.

### 2.2 Interactions between single NRs.

The NR composites discussed above were random collections of NRs and NPs suspended in a solution or built into a polymer matrix. The optical density of such composites was calculated as a linear superposition of extinctions of individual components. This approach assumes that the elements, NRs and NPs, are weakly interacting. We have shown numerically that the interactions can be neglected if the inter-nanocrystal distance in a composite is about or larger than the sizes of components. Supporting information gives a few examples of interactions between nanocrystals. If however the inter-nanocrystal distance is smaller than the lengths of components, the electromagnetic interactions become strong and most likely destroy the window effect. In Figure S6, we show such case. In the model used in Figure S6, we arrange 17 NRs in an array with a relatively small inter-NR distance. Then, we calculate the total extinction and compare it with the case of linear superposition. Whereas the linear superposition has a transparency window at ~ *1500nm*, the metamaterial structure, which looks like a pipe organ, does not exhibit the optical window at all. In this structure, strong inter-NR electromagnetic interactions modify the whole spectrum and completely destroy the window effect.

Another interesting question is whether the interactions between single nanocrystals can help to create a more narrow transparency window. In a nanostructure composed of two NRs, we expect the appearance of the splitting and the related double-peak spectral structure. The question is whether such double peak structure can be used for the transparency window effect. Figure 8 shows that the simple NR-NR interaction does not improve the window effect. The window effect of the two interacting NRs is almost identical to the effect created by two independent NRs with appropriately-chosen sizes. We should note that this is just one particular case of inter-nanocrystal interaction and there may be other cases when the interactions can be helpful for the transparency window effects.

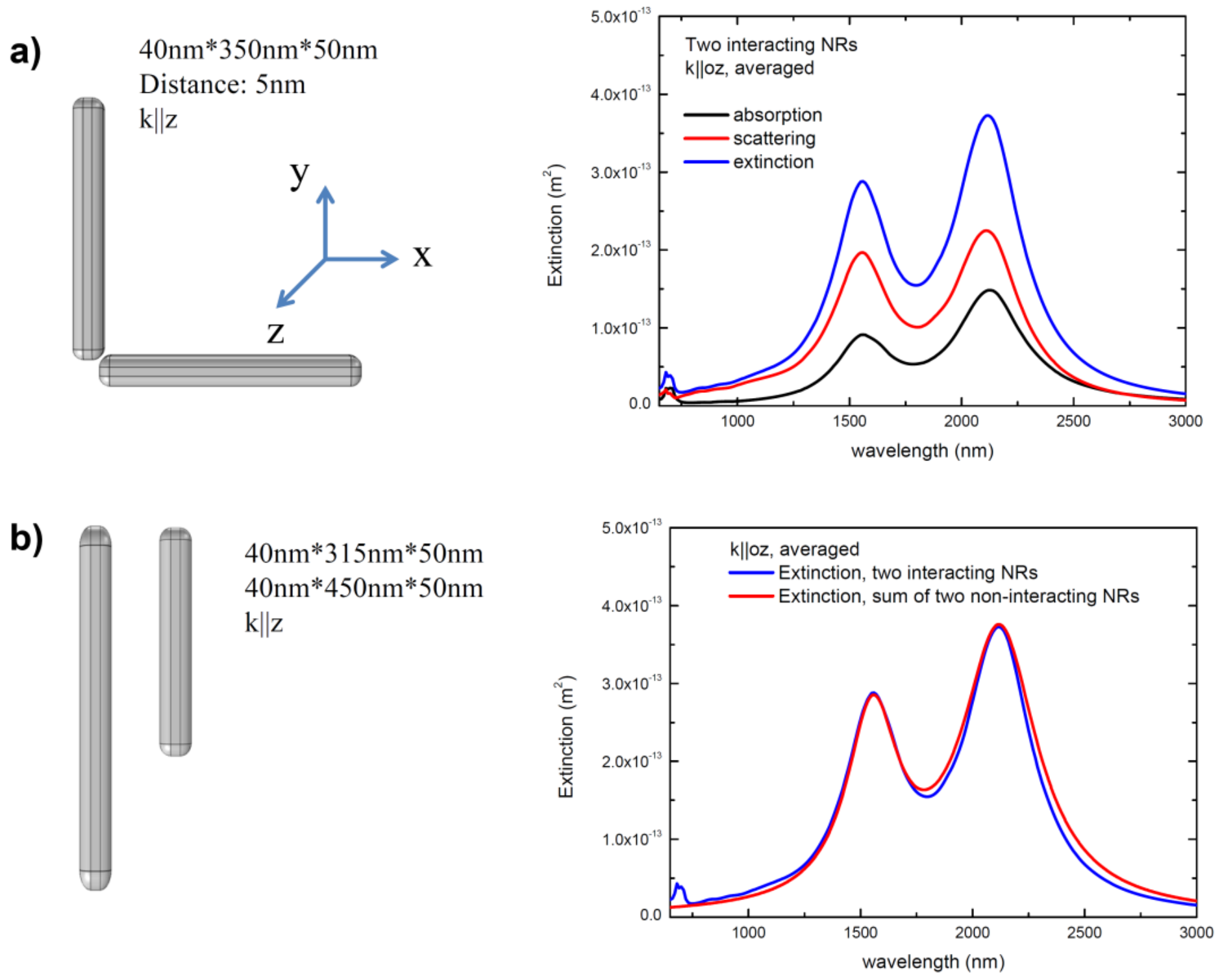

**Figure 8:** (a) Extinctions of a double-NR structure; single NRs in this structure are strongly interacting and the resultant spectrum exhibits a plasmon splitting. (b) Comparison of extinctions for the double-NR structure and the pair of non-interacting NRs. The extinction of the NR pair can be well approximated as a simple sum of extinctions of single NRs with the appropriately-chosen lengths. Extinctions were averaged over the polarization of light assuming the normal incidence. Insets show the models.

## 3. Metastructures and metamaterials.

We now look at more complex structures. The idea is to find more complex elementary blocks that can be more compact and convenient. Such nanocomposite material can occupy less space and also have fewer elements which may simplify a fabrication. We could already see that two interacting NRs do not give much advantage since the window effect is similar to the case of two non-interacting nanocrystals (Fig. 8). Additionally, the structure with two perpendicular NRs exhibits an asymmetric double peak spectrum whereas, for a material with a transparency window, we better have elements with symmetric double-peak spectra. For this, we will try now simple and multi-bar nano-crosses.

### 3.1 Nano-crosses.

We aim to find an element that has a symmetric double-peak structure which can be used for a metamaterial with a transparency window. However, simple nano-crosses give typically asymmetric windows (Fig. 9). Therefore we can try a cross with two perpendicular bars. After the optimization of the parameters, we show that the two-bar cross can produce a symmetric window. A symmetric window effect appears only for certain, specially-chosen geometrical parameters. For example, the two-bar cross with a large inter-bar distance in Fig. 10 produces an asymmetric window since the long-wavelength peak becomes enhanced. The next step is to use nano-crosses as elements in a metastructure.

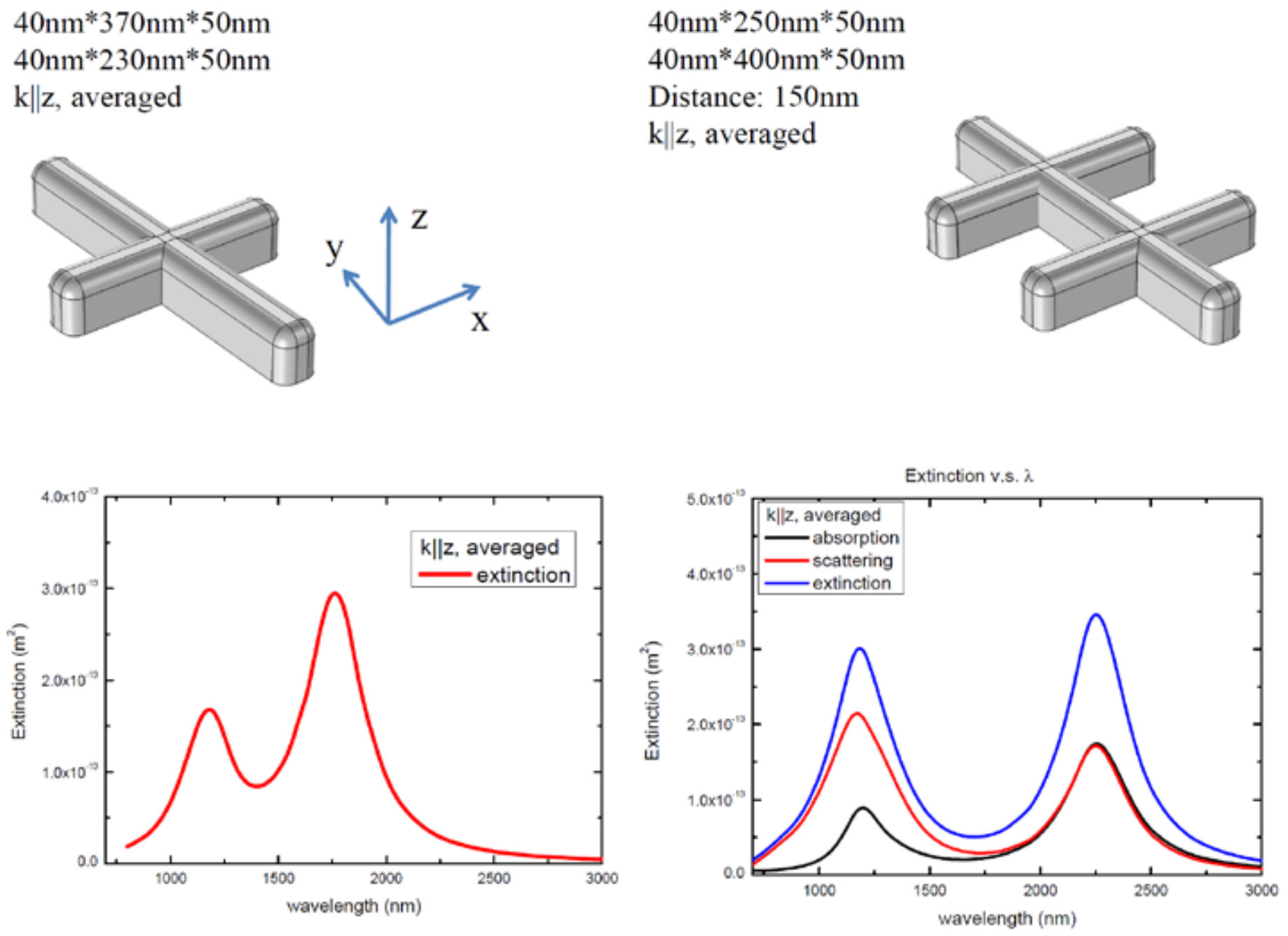

**Figure 9:** Extinction spectra of simple and two-bar nano-crosses. The two-bar nano-cross shows an almost symmetric double-peak structure that is needed for the window effect. The extinctions are calculated for the normal incidence and then averaged over the polarization of light.

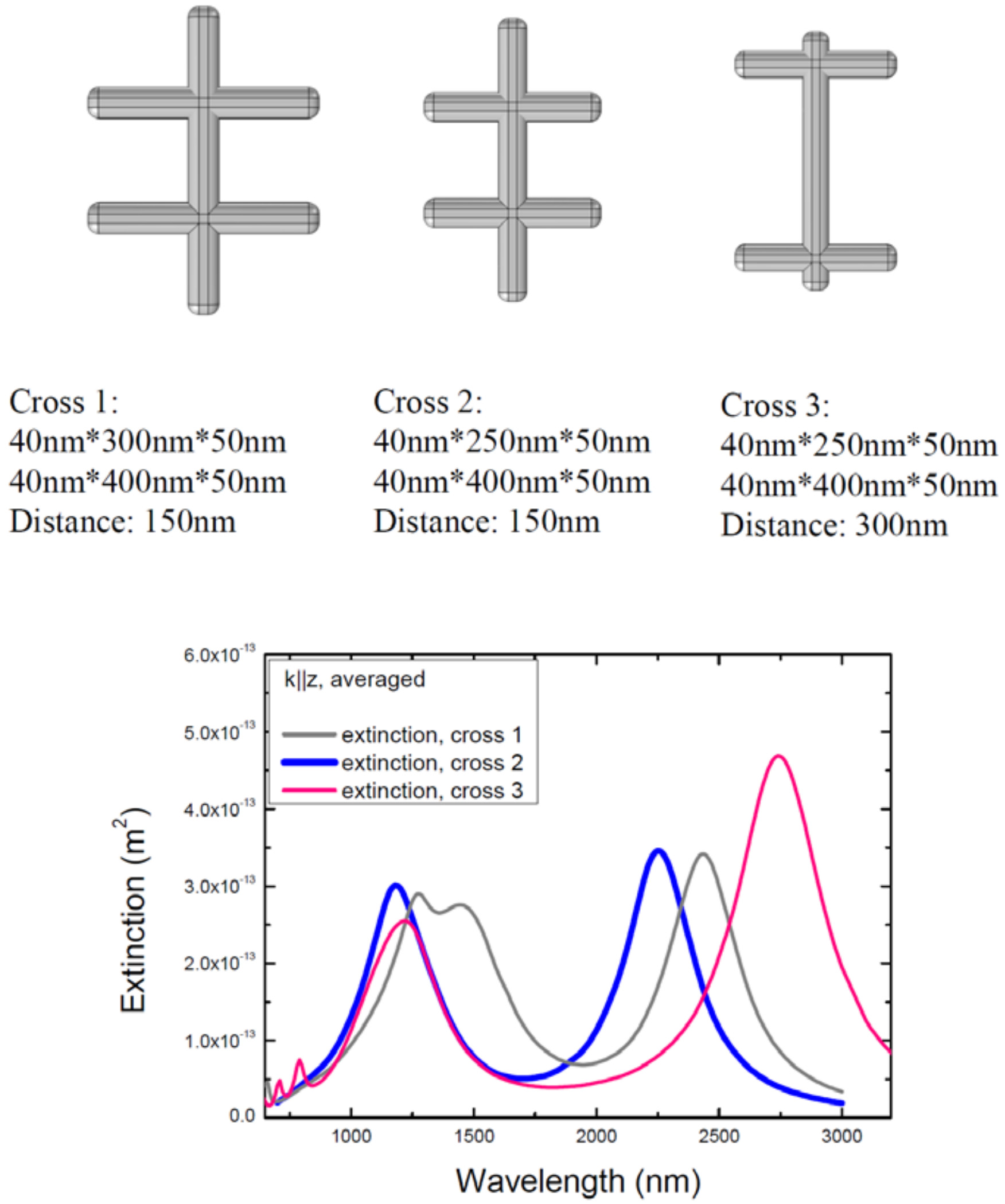

**Figure 10:** Extinction spectra of two-bar nano-crosses with different parameters. The two-bar nano-cross in the center exhibits an almost symmetric window effect. Again, the extinctions are calculated for the normal incidence and then averaged over the polarization of light.

### 3.2 Composition of nano-crosses, nanorods and nano-disks.

We now involve three crosses, two NRs, and two nano-disks (NDs). Figure 11 shows the composition. All these elements can be fabricated with the lithography technologies. The height of the nanostructures is taken as $h = 50nm$ and the width of the bars in these structures is chosen as $w = 40nm$. Three nano-crosses give three double-peak structures with the transparency window in the middle. NRs and NDs block the transmission in the short- and long-wavelength intervals. The resulting spectrum exhibits strong extinction in the interval *400nm-4.5μm* and a transparency window centered at *1660nm*.

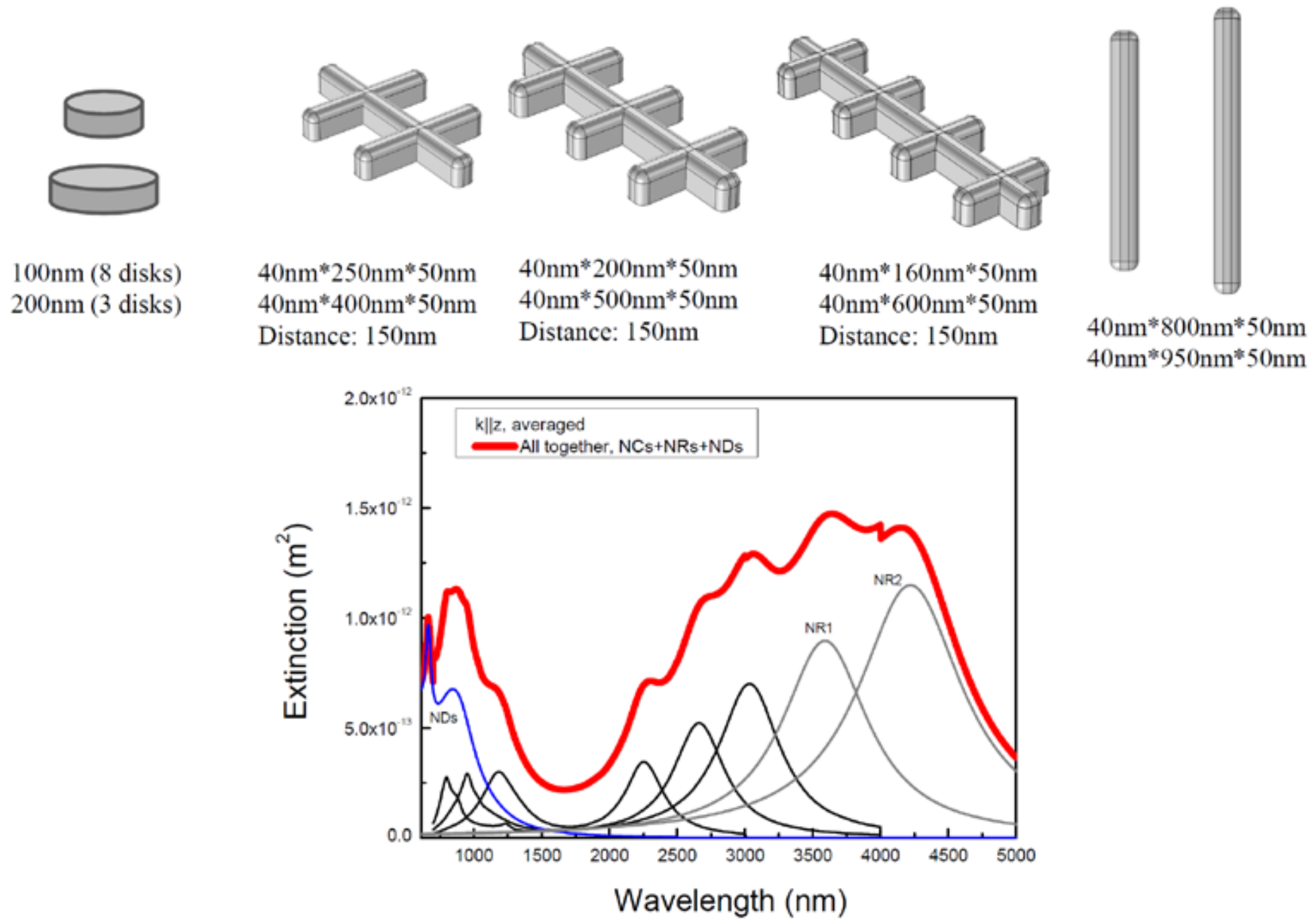

**Figure 11:** Extinction spectrum of the composition incorporating three nano-crosses, two nanorods and two nano-disks. All elements are made of gold. Insets show the geometries of the elements.

The optical medium with the transparency window can also be fabricated as a planar meta-material on a substrate. Fig. 12 shows this case. The distances between the nanocrystals in this metastructure are long enough to avoid interference between the elements. The distances can be found in Figure S9. In addition, in this metastructure, we like to suppress the interference effects by orienting the nanocrystals. To do this, we arrange the elements in the way to avoid quasi-resonant interactions between the elements. For example, we orient the two-bar and three-bar crosses perpendicular to each other. Similarly, we make the three-bar and four-bar crosses perpendicular as well. Therefore the three interacting nano-crosses in our structure do not show prominent effects of interaction. Also, two long NRs are placed at a large distance from each other and on the common line. The nano-disks are well separated and do not interact much. We confirm that the interactions in our structure are weak by direct numerical COMSOL-based calculations shown in Supporting Information. By positioning and orienting the elements in the certain way, we can avoid strong interactions between nanocrystals and distortions of the window effect.

We have perfumed a number of calculations of interacting nanocrystals and found the following simple rules to construct metastructures with weak interactions. (1) We orient two quasi-resonant elongated elements (NRs or nano-crosses) perpendicular to each other. Then, the L-resonances do not interact due to the symmetry. (2) Long NRs with very strong plasmon resonances better be placed on a common line. (3) We place the elements that are resonant or quasi-resonant as far as possible from each other on a substrate. (4) Spherical and cylindrical elements should have inter-particle distances that are about their sizes. The logistics behind these rules is to reduce the interference effects in a metastructure using the geometrical approach.

We also should make a few final comments on fabrication methods for the described metastructures. Thin substrates (glass or polymer) with metastructures can be randomly dispersed in a medium (liquid, polymer or air) or oriented as parallel multi-layers in a polymer matrix (Fig. 12). The resulting transmission spectrum of such system shows a strong transparency window effect (Fig. 12). This is another possibility to realize an optical material with both a strong broadband extinction spectrum and a narrow transmission window. Since the lithography allows us to fabricate nanorods and nanowires with an almost arbitrary length (of course, within the reasonable limits ☺), the proposed optical metamaterials may have transparency windows in the wavelength interval from *600nm* to *1mm*.

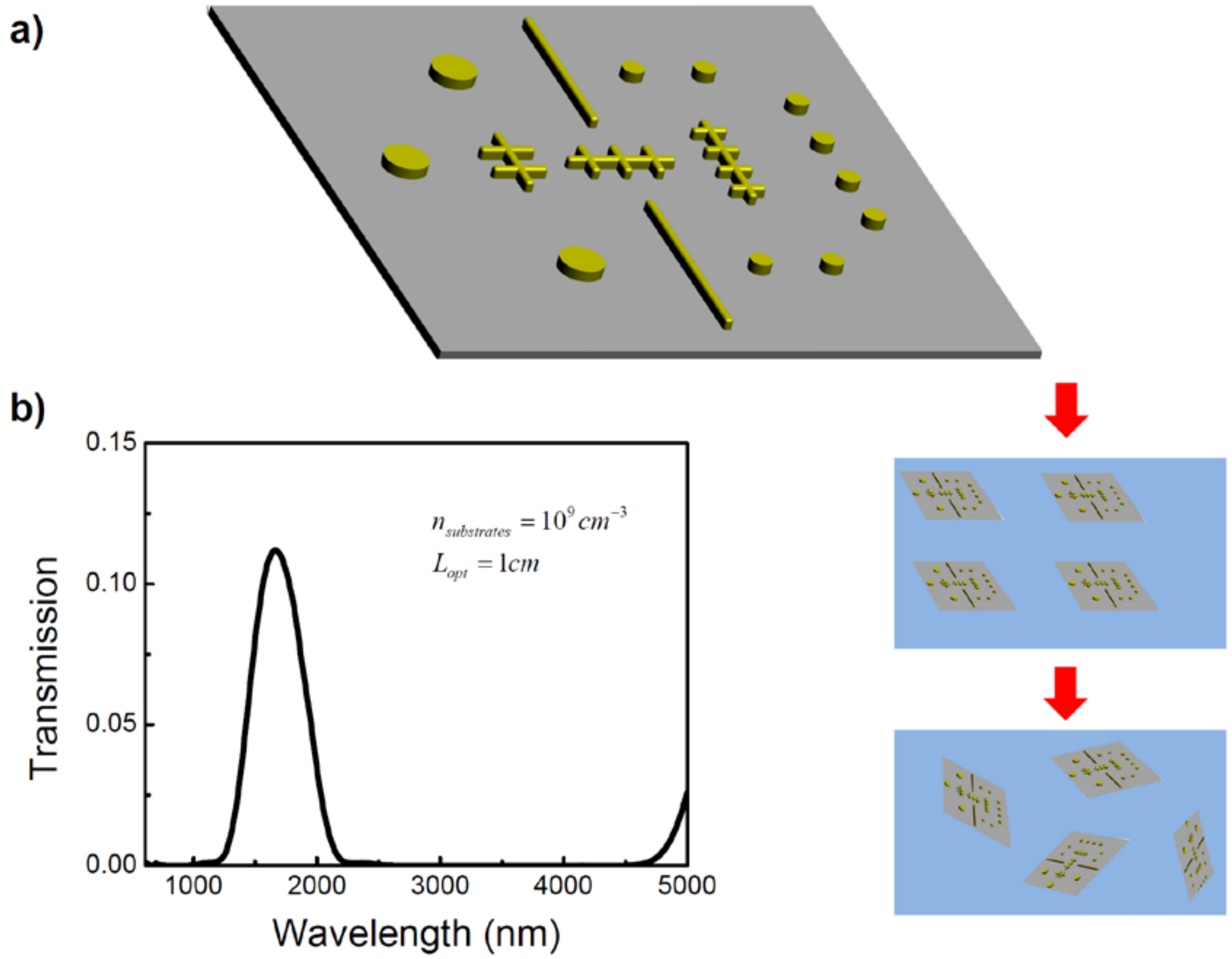

**Figure 12:** **a)** Model of planar plasmonic metastructure with a transparency window. **b)** Calculated transmission spectrum of a solution with despised metastructures having random orientations. Insets show two models: A layered medium based on a polymer matrix and a liquid solution with dispersed metastructures.

**Conclusions**

We have proposed and described a novel class of nanostructured materials with very peculiar optical properties. This nanomaterial strongly attenuates light in a very broad spectral interval, but exhibits a narrow window of transmission. The principle to construct these optical media is based on narrow plasmonic peaks of metal nanorods and nano-crosses. The optical medium is constructed as a mixture of elements with a specially-designed distribution of sizes. Nanomaterials with a transparency window can be fabricated by using both the colloidal synthesis and the lithography technologies. These materials can be made as solutions of single nanocrystals or as a two-dimensional metamaterial on a substrate. Optical media with a transparency window, which can be made at a chosen wavelength, can be in the forms of liquid solutions, polymer films, coatings and aerosols. Using plasmonic nanorods and nanocrystals, the transmission window can be designed at any particular wavelength in the interval from *400nm* to tens of microns and further to the infrared.

**Acknowledgements**

We thank Z. Fan for help with the computations. H. Zhang and A. O. Govorov were supported by the U.S. Army Research Office under grant number W911NF-12-1-0407 and by the Volkswagen Foundation (Germany). Use of the Computing Cluster at the Center for Nanoscale Materials was supported by the U. S. Department of Energy, Office of Science, Office of Basic Energy Sciences, under Contract No. DE-AC02-06CH11357. H.V.D. gratefully acknowledges support from NRF-RF-2009-09 and NRF-CRP-6-2010-2 as well as TUBA and ESF EURYI.

# Supporting Information for

## “Plasmonic Metamaterials and Nanocomposites with

## a Narrow Transparency Window in a Broad Extinction Spectrum”

*Hui Zhang, Hilmi Volkan Demir, and Alexander O. Govorov**

**Figure S1:** Absorption of the components in the discrete NR nanocomposite material.

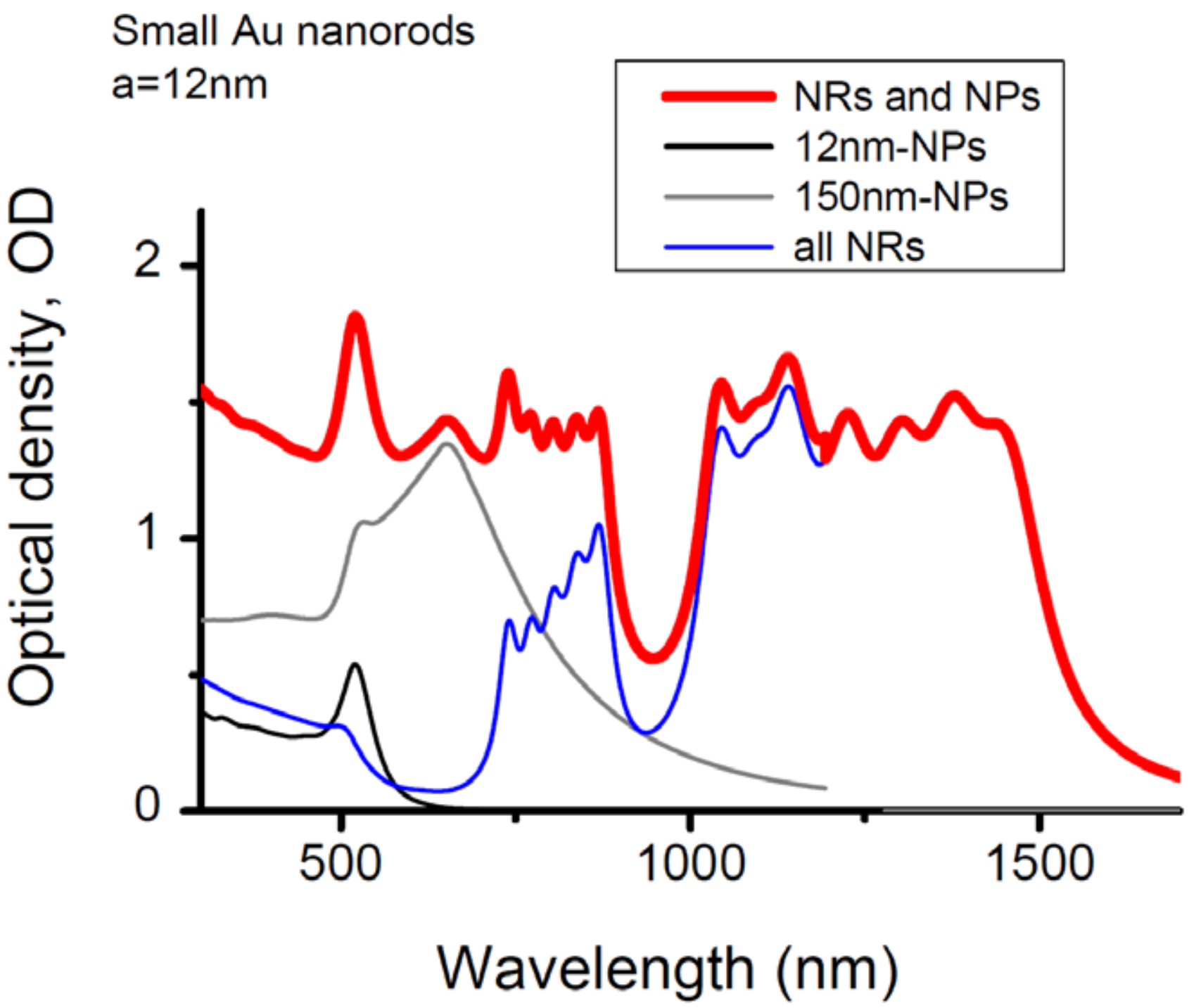

$c_0 = 0.5 \cdot 10^{-10} M$

$n_0 = 3 \cdot 10^{10} cm^{-3}$

$L_{opt} = 1cm$

$n_{12\text{nm-NP}} = 45 \cdot n_0$

$n_{150\text{nm-NP}} = 0.1 \cdot n_0$

$n_{44\text{nm-NR}} = 1.5 \cdot n_0$

$n_{48\text{nm-NR}} = 1.0 \cdot n_0$

$n_{52\text{nm-NR}} = 1.0 \cdot n_0$

$n_{56\text{nm-NR}} = 1. \cdot n_0$

$n_{60\text{nm-NR}} = 1.3 \cdot n_0$

$n_{80\text{nm-NR}} = 1.2 \cdot n_0$

$n_{86\text{nm-NR}} = 0.6 \cdot n_0$

$n_{92\text{nm-NR}} = 1.0 \cdot n_0$

$n_{102\text{nm-NR}} = 0.8 \cdot n_0$

$n_{112\text{nm-NR}} = 0.6 \cdot n_0$

$n_{122\text{nm-NR}} = 0.6 \cdot n_0$

$n_{132\text{nm-NR}} = 0.6 \cdot n_0$

**Figure S2:** Discrete hybrid composite with Au NRs, Au NPs and CdTe QDs.

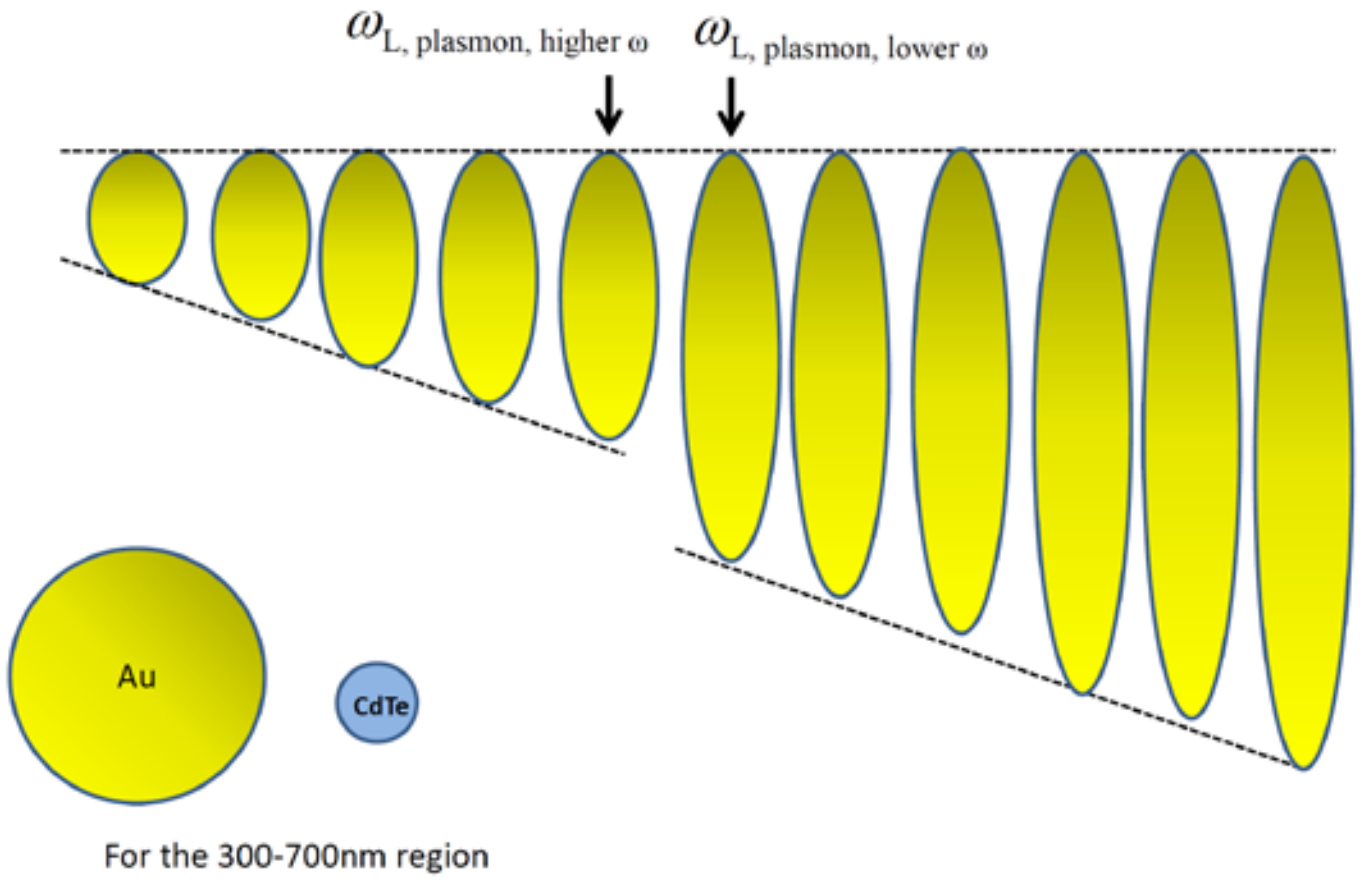

**Figure S3:** Optical density of hybrid composite with Au NRs, Au NPs and CdTe QDs.

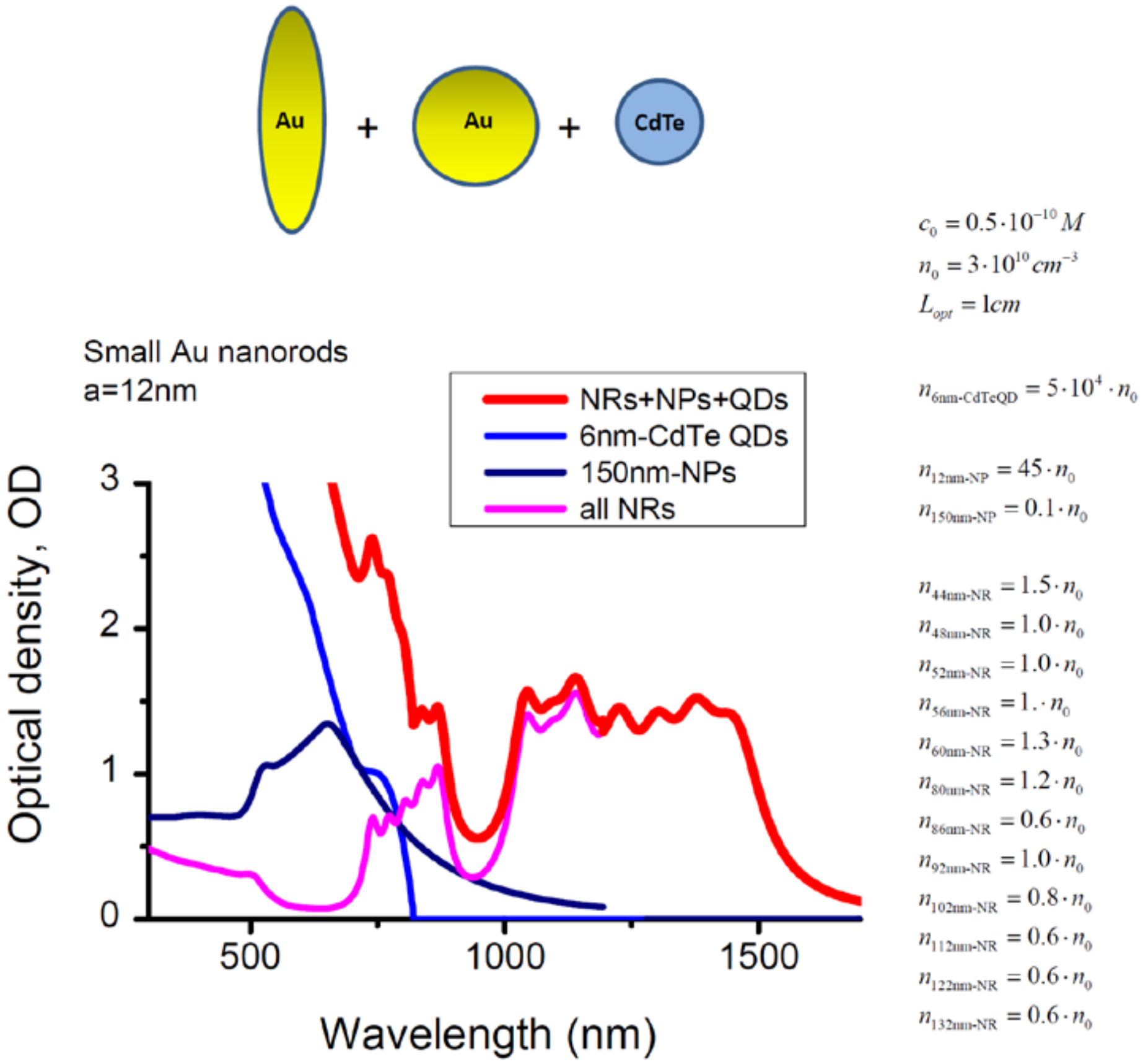

**Figure S4:** Excitations of large lithographic Au-nanorods calculated by COMSOL.

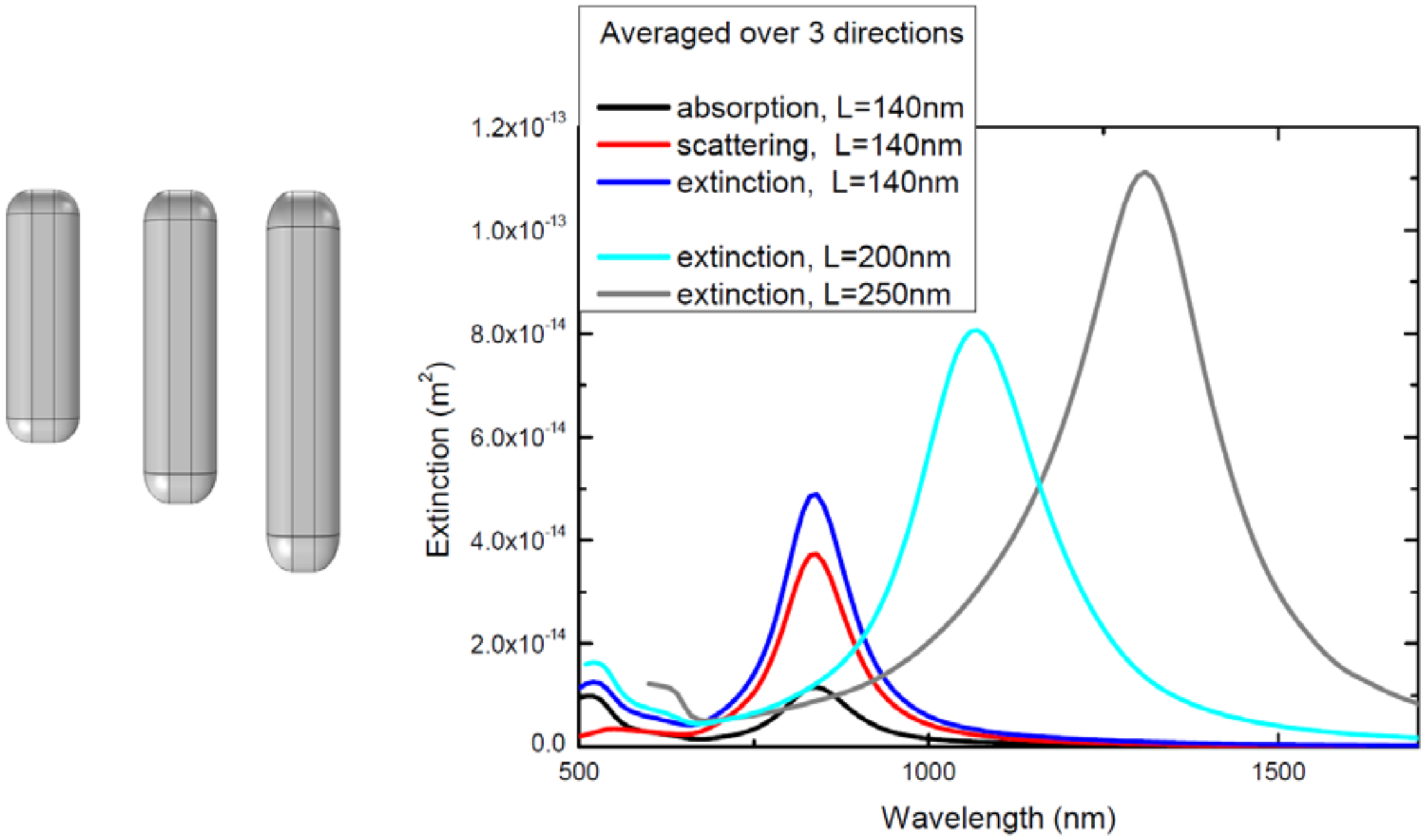

**Figure S5: Upper graph:** Optical density of a collection of large NRs and NPs. 100nm-NPs give a strong contribution to absorption in the short-wavelength interval. **Lower graph:** Normalized transmission spectra for various optical paths. The window shrinks with increasing optical path.

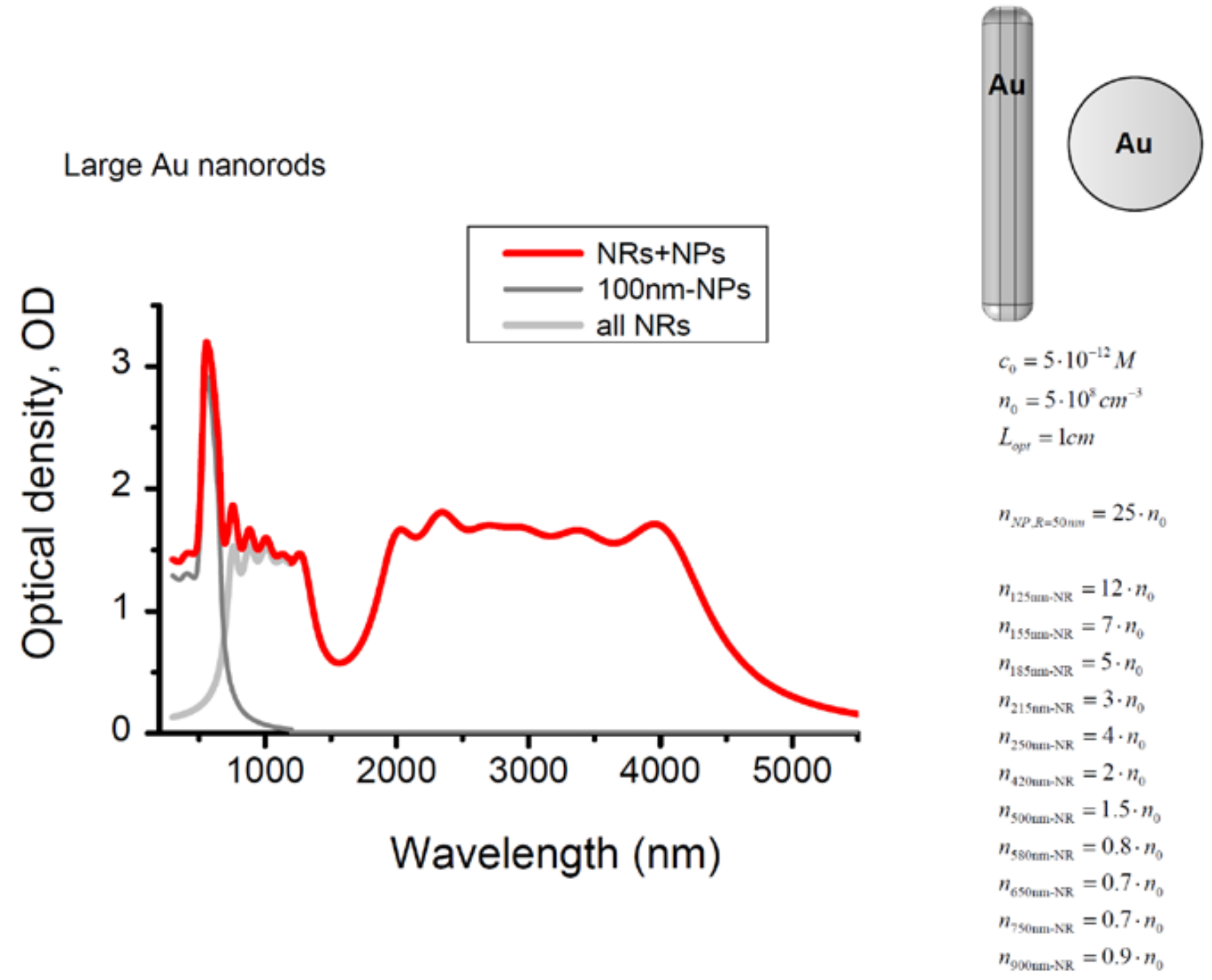

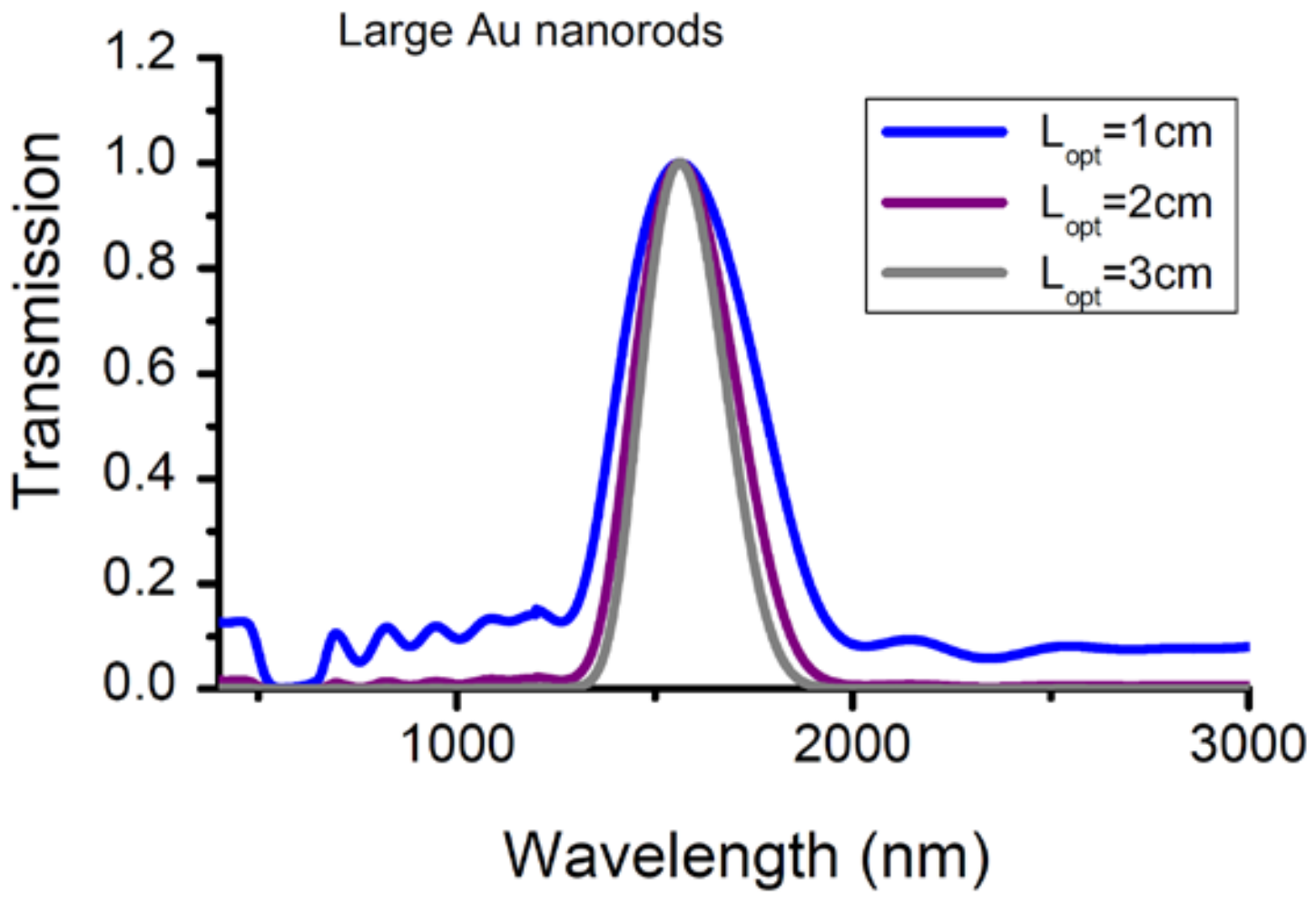

**Figure S6:** Model and optical properties of a plasmonic metastructure composed of NRs. The lower graph shows the extinctions of meta-structure and linear superposition of NRs. We can see that the interaction in the meta-structure have destroyed the optical window.

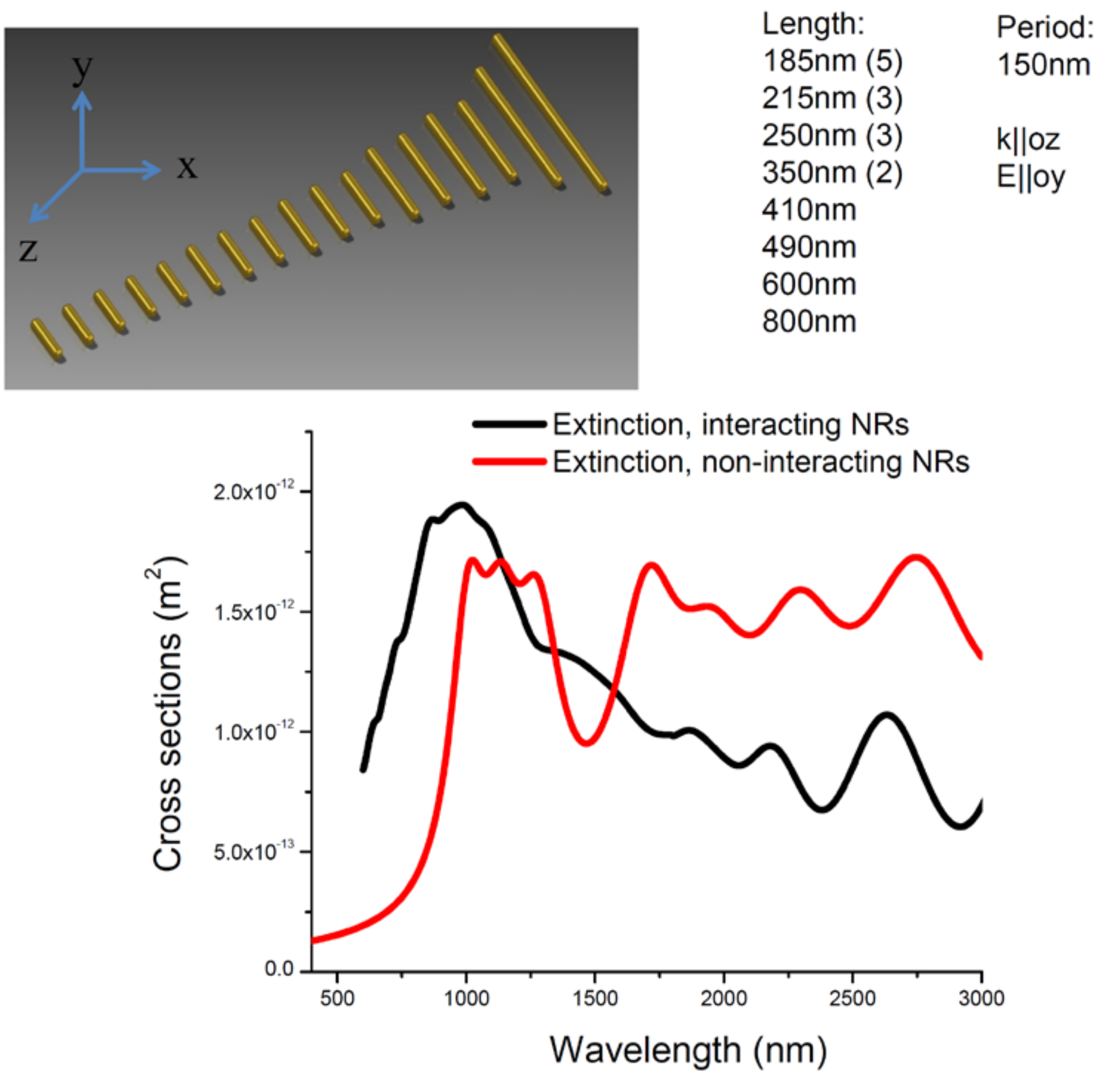

**Figure S7:** Interactions between single NRs arranged in pairs and placed at difference distances. We see that effect of the interaction becomes small when the inter-NR distance is about the length of NR. In additional, the mutual orientation of NRs is very important and can be used to minimize the interaction. For two perpendicular NRs (right panel), the interaction is very weak, while for two parallel NRs (left panel) the interaction remains noticeable at relatively large distances.

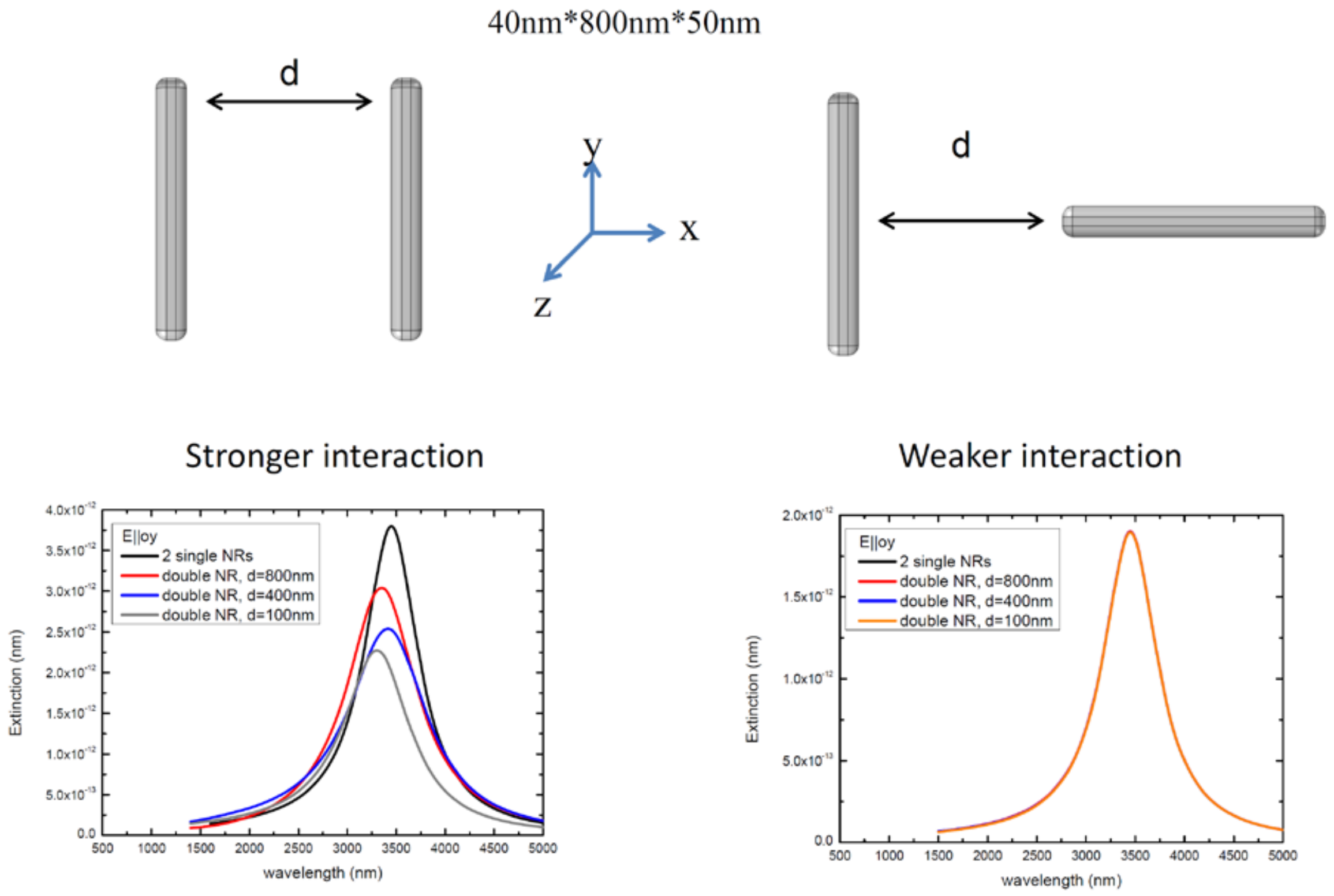

**Figure S8:** This graph shows the distances between the elements in the metastructure investigated in the main text and shown in Fig. 12.

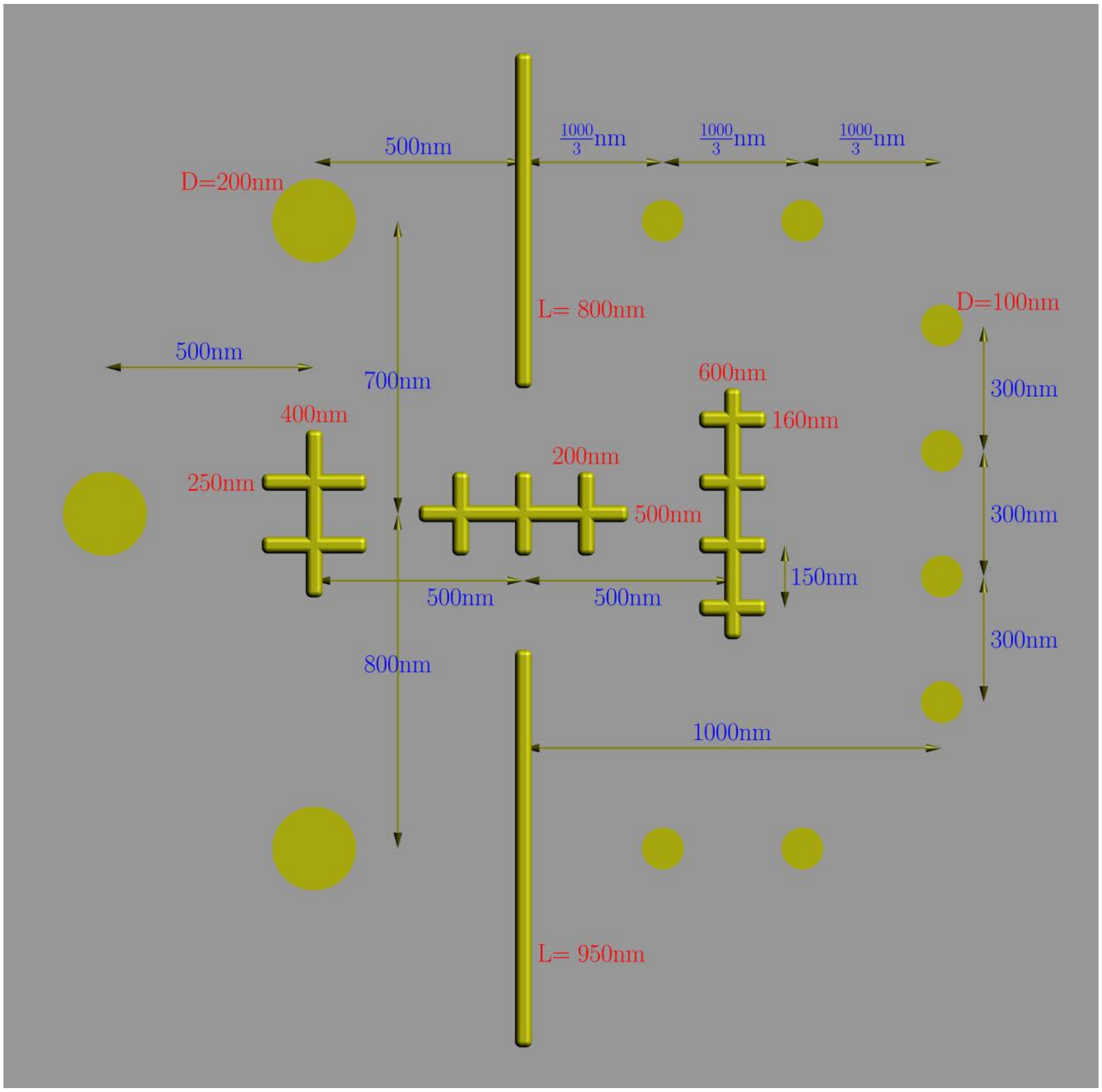

.

**Figure S9:** Interactions between nanocrystals in the metastructure. We assume the geometry and the inter-nanocrystal distances shown in Figures 12 and S8. We see that the electromagnetic interactions between the nanocrystals do not change noticeably the window effect in the metastructure. We designed this metastructure intentionally to reduce the interactions between the units since these interactions can disturb or even destroy the window effect. The simple rules to construct the structures with weak interactions are discussed in the main text.

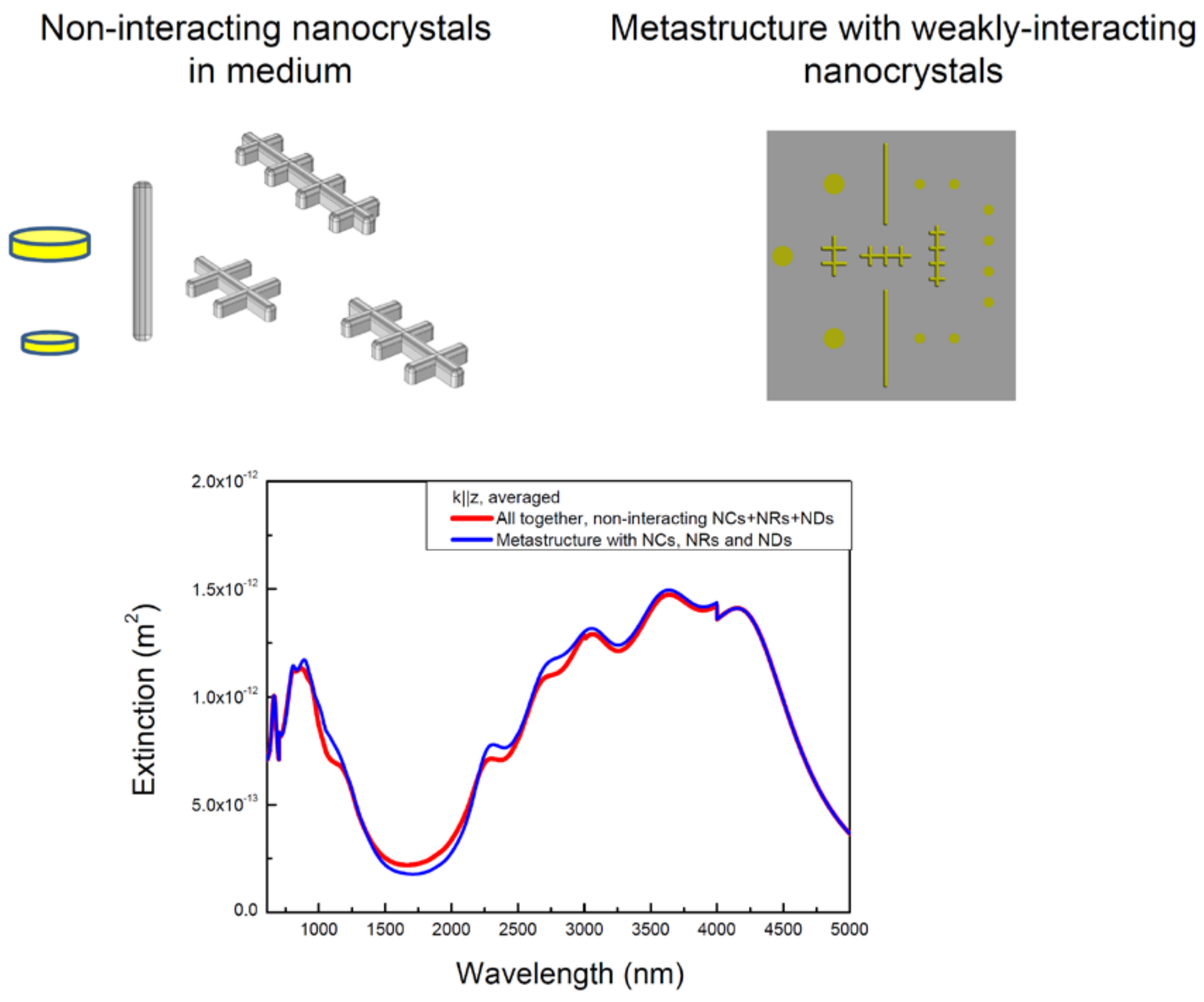

[1]Maier, S. A. *Plasmonics: Fundamentals and applications*, Springer: N.Y., 2007.

[2] Novotny, L.; Hecht, B. *Principles of Nano-Optics*, Cambridge Univ. Press, Cambridge, UK, 2006.

[3] *Complex-shaped Metal Nanoparticles. Bottom-Up Syntheses and Applications*, Sau, T. K.; Rogach, A. L., Eds.; Wiley-VCH, Weinheim, 2013.

[4] Link, S.; El-Sayed, M. A. Spectral Properties and Relaxation Dynamics of Surface Plasmon Electronic Oscillations in Gold and Silver Nanodots and Nanorods. *The Journal of Phys. Chem. B* **1999,** *103*, 8410-8426.

[5] Dondapati, S. K.; Sau, T. K.; Hrelescu, C.; Klar, T. A.; Stefani, F. D.; Feldmann, J. Label-free Biosensing Based on Single Gold Nanostars as Plasmonic Transducers. *ACS Nano* **2010** *4* (11), 6318-6322.

[6] Shelby, R. A.; Smith, D. R.; Schultz, S. Experimental Verification of a Negative Index of Refraction, *Science* **2001**, *292*, 77-79.

[7] Luk'yanchuk, B.; Zheludev, N. I.; Maier, S. A.; Halas, N. J.; Nordlander, P.; Giessen, H.; Chong, C. T. The Fano resonance in plasmonic nanostructures and metamaterials, *Nature Mat.* **2001**, *9*, 707–715.

[8] Fan, J. A.; Wu, C.; Bao K.; Bao J.; Bardhan R.; Halas N. J.; Manoharan V. N.; Nordlander P.; Shvets, G.; Capasso F. Self-assembled plasmonic nanoparticle clusters. *Science* **2010**, *328*, 1135–1138.

[9] Verellen, N.; Sonnefraud, Y.; Sobhani, H.; Hao, F.; Moshchalkov, V. V.; Dorpe, P. V.; Nordlander, P.; Maier, S. A. Fano Resonances in Individual Coherent Plasmonic Nanocavities, *Nano Lett.*, **2009**, *9*, 1663-1667.

[10] Shvets, G.; Urzhumov, Y. A. Engineering the electromagnetic properties of periodic nanostructures using electrostatic resonances. *Phys. Rev. Lett.* **2004**, *93*, 243902, 1-4.

[11] Wiederrecht, G. P.; Wurtz, G. A.; Hranisavljevic, J. Coherent coupling of molecular excitons to electronic polarizations of noble metal nanoparticles, *Nano Letters,* **2004**, *4*, 2121–2125.

[12] Zhang, W.; Govorov, A. O.; Bryant, G. W. Semiconductor-metal nanoparticle molecules: hybrid excitons and non-linear Fano effect, *Phys. Rev. Lett.* **2006**, *97,* 146804, 1-4.

[13] Shah, R. A.; Scherer, N. F; Pelton, M.; Gray, S. K. Ultrafast reversal of a Fano resonance in a plasmon-exciton system, *Phys. Rev. B* **2013**, *88*, 075411, 1-7.

[14] Zhang, J.; Tang, Y.; Lee, K.; Ouyang, M. Tailoring light-matter-spin interactions in colloidal hetero-nanostructures. *Nature* **2010**, *466*, 91-95.

[15] Zhang, S.; Genov, D. A.; Wang, Y.; Liu, M.; Zhang, X. Plasmon-Induced Transparency in Metamaterials, *Phys. Rev. Lett.* **2008**, *101*, 047401, 1-4.

[16] Liu, N.; Langguth, L.; Weiss, T.; Kästel, J.; Fleischhauer, M.; Pfau, T.; Giessen, H. Plasmonic analogue of electromagnetically induced transparency at the Drude damping limit, *Nature Mat.* **2009**, *8*, 758 - 762.

[17] Biswas, S.; Duan, J.; Nepal, D.; Park, K.; Pachter, R.; Vaia, R. Plasmon induced transparency in the visible wavelengths enabled by self-assembled metamaterials, *Nano Letters*, **2013**, *13*, 6287-6291.

[18] Landau L. D.; Lifshitz, E. M. *Electrodynamics of Continuous Media.* Pergamon, N. Y., 1960.

[19] van de Hulst, H. C. *Light Scattering by Small Particles*, Dover Publications, N.Y., 1981.

[20] Johnson, P. B.; Christy, R. W. Optical Constants of the Noble Metals. *Phys. Rev. B* **1972**, *6*, 4370-4379.

[21] Vial, A.; Grimault, A.-S.; Macías, D.; Barchiesi, D.; de La Chapelle, M. L. Improved analytical fit of gold dispersion: Application to the modeling of extinction spectra with a finite-difference time-domain method. *Phys. Rev. B* **2005**, *71*, 085416, 1-7.

[22] http://icb.u-bourgogne.fr/nano/manapi/saviot/mie/index.en.html

[23] http://www.nanopartz.com/bare_gold_nanorods.asp

[24] Lee, J.; Hasan, W.; Stender, C. L.; Odom, T. W. Pyramids: A Platform for Designing Multifunctional Plasmonic Particles, *Acc. Chem. Res.*, **2008**, *41*, 1762-1771.

[25] Mark, A. G.; Gibbs, J. G.; Lee T.-C.; Fischer, P. Hybrid nanocolloids with programmed three-dimensional shape and material composition, *Nature Mat.* **2013**, *12*, 802–807.

[26] Yeom, B.; Zhang, H.; Zhang, H.; Park, J.; Kim, K.; Govorov, A. O.; Kotov, N. A. Chiral Plasmonic Nanostructures on Achiral Nanopillars, *Nano Lett.* **2013**, *13*, 5277–5283.